\documentclass[10pt, sigconf]{acmart}

\newcommand{\chancery}[1]{{\fontfamily{pzc}\selectfont #1}}
\newcommand{\chancerybold}[1]{{\fontfamily{pzc}\selectfont\textbf{#1}}}

\usepackage{blindtext}
\usepackage[utf8]{inputenc}
\usepackage{epsfig,xspace,url,xcolor}
\usepackage{subcaption}
\usepackage{thmtools}
\usepackage{thm-restate}
\usepackage{mathtools}
\usepackage{tikz}
\usepackage{caption}
\usepackage{graphicx}
\usepackage[ruled, lined, linesnumbered, commentsnumbered, longend]{algorithm2e}
\usepackage[noend]{algpseudocode}
\usepackage{url}
\usepackage{colortbl}
\usepackage{xcolor}

\usepackage[subtle]{savetrees}

\usepackage{wrapfig}

\newcommand{\name}{{\chancery{Vermilion}}\xspace}
\newcommand{\namebold}{{\chancerybold{Vermilion}}\xspace}

\renewcommand\footnotetextcopyrightpermission[1]{} 
\input{macros}

\let\ACMmaketitle=\maketitle
\renewcommand{\maketitle}{\begingroup\let\footnote=\thanks \ACMmaketitle\endgroup}

\makeatletter
\def\author@bx@sep{0pc}
\makeatother

\begin{document}

\title{Vermilion: A Traffic-Aware Reconfigurable Optical Interconnect with Formal Throughput Guarantees}

\author{Vamsi Addanki}
\affiliation{
  \institution{TU Berlin}
}
\author{Chen Avin}
\affiliation{
  \institution{Ben-Gurion University of the Negev}
}

\author{Goran Dario Knabe}
\affiliation{
  \institution{TU Berlin}
}

\author{Giannis Patronas}
\affiliation{
  \institution{NVIDIA}
}

\author{Dimitris Syrivelis}
\affiliation{
  \institution{NVIDIA}
}

\author{Nikos Terzenidis}
\affiliation{
  \institution{NVIDIA}
}

\author{Paraskevas Bakopoulos}
\affiliation{
  \institution{NVIDIA}
}

\author{Ilias Marinos}
\affiliation{
  \institution{NVIDIA}
}

\author{Stefan Schmid}
\affiliation{
  \institution{TU Berlin}
}

\renewcommand{\shortauthors}{Addanki et al.}

\sloppy
\begin{abstract}
The increasing gap between datacenter traffic volume and the capacity of electrical switches has driven the development of reconfigurable network designs utilizing optical circuit switching. Recent advancements, particularly those featuring periodic fixed-duration reconfigurations, have achieved practical end-to-end delays of just a few microseconds. However, current designs rely on multi-hop routing to enhance utilization, which can lead to a significant reduction in worst-case throughput and added overhead from congestion control and routing complexity. These factors pose significant operational challenges for the large-scale deployment of these technologies.

We present \name, a reconfigurable optical interconnect that breaks the throughput barrier of existing periodic reconfigurable networks, without the need for multi-hop routing --- thus eliminating congestion control and simplifying routing to direct communication. 
\name adopts a traffic-aware approach while retaining the simplicity of periodic fixed-duration reconfigurations, similar to RotorNet. We formally establish throughput bounds for \name, demonstrating that it achieves at least $33\%$ more throughput in the worst-case compared to existing designs.
The key innovation of \name is its \emph{short} traffic-aware periodic schedule, derived using a matrix rounding technique. This schedule is then combined with a traffic-oblivious periodic schedule to efficiently manage any residual traffic.
Our evaluation results support our theoretical findings, revealing significant performance gains for datacenter workloads.

\end{abstract}

\maketitle
\thispagestyle{plain}
\pagestyle{plain}

\section{Introduction}
\label{sec:introduction}

\begin{figure}[t]
    \centering
    \vspace*{3mm}
    \includegraphics[width=1\linewidth]{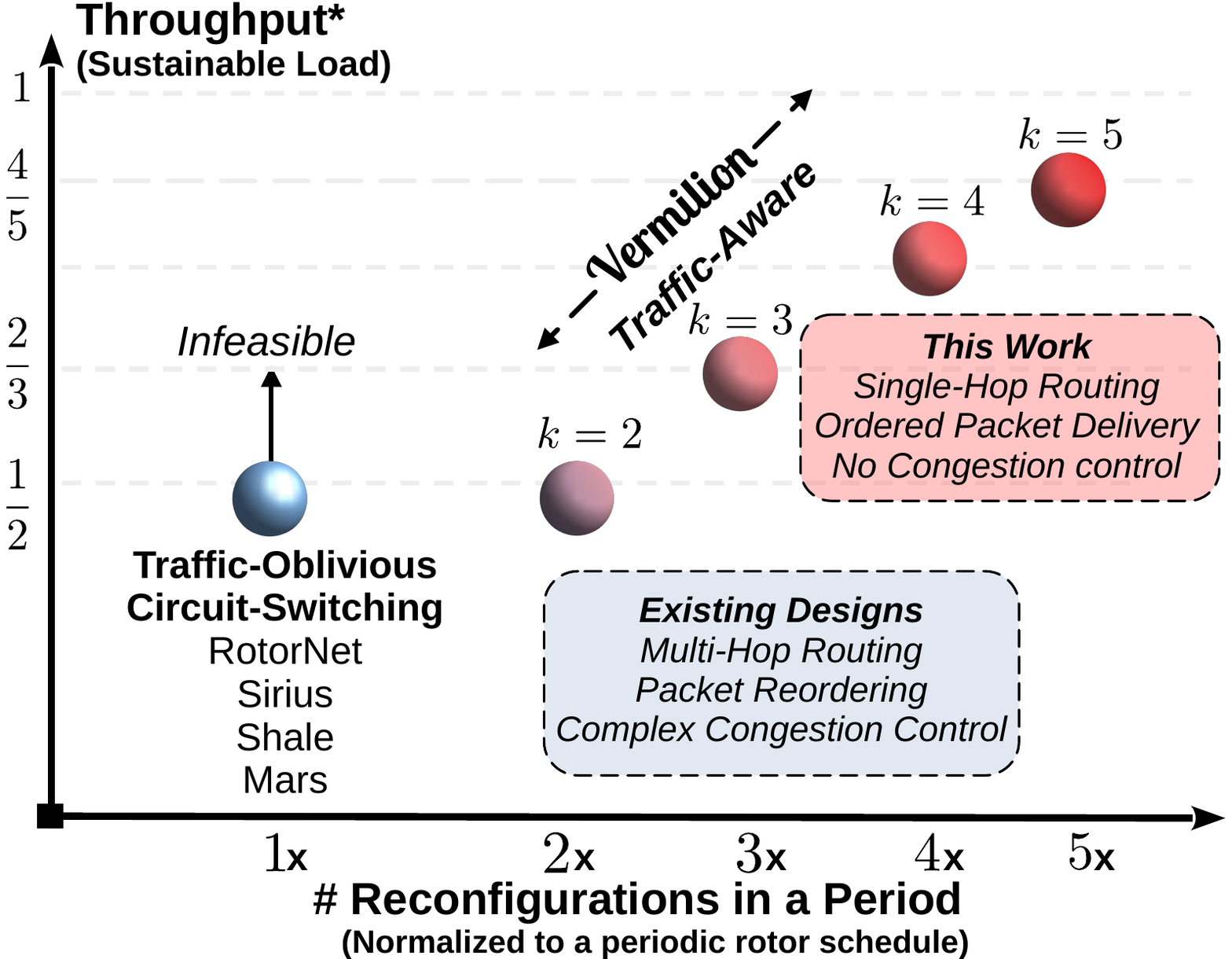}
    \caption{Existing designs based on periodic optical circuit-switching are oblivious to traffic patterns, requiring complex multi-hop routing and congestion control, which reduces throughput. \name overcomes this limitation by introducing a few additional fixed-duration reconfigurations per period in a traffic-aware manner, while significantly simplifying both routing and congestion control.
    }
    \vspace{-4mm}
    \label{fig:intro}
\end{figure}

Datacenters have experienced explosive growth in overall network traffic volume over the past decade~\cite{10.1145/2785956.2787508}.
With the recent introduction of high-bandwidth Machine Learning workloads into datacenters, the peak network traffic is expected to increase even more rapidly~\cite{10.1145/3544216.3544265}.
Unfortunately, traditional networks, which are built using electrical packet switches, struggle to keep up with this growing demand~\cite{10.1145/3387514.3406221}. Further, the rapid evolution of datacenter applications and their changing bandwidth requirements implies: ``the best laid plans quickly become outdated and inefficient, making incremental and adaptive evolution a necessity''~\cite{10.1145/3544216.3544265}.
This led to the emergence of novel technologies based on reconfigurable optical circuit switches~\cite{10.1145/3098822.3098838,10.1145/3387514.3406221,10.1145/2934872.2934911,10.1145/1851182.1851223,10.1145/3544216.3544265}.
Two prominent types of reconfigurable datacenter networks have emerged recently: traffic-oblivious~\cite{10.1145/3098822.3098838,10.1145/3387514.3406221,10.1145/3579312,10.1145/3651890.3672248} and traffic-aware~\cite{10.1145/2934872.2934911,10.1145/1851182.1851223,10.1145/3579449} networks. These networks are optically circuit-switched and feature bufferless switches. The circuits can be reconfigured, enabling the network topology to adapt dynamically to evolving communication patterns in datacenter workloads, which can potentially improve performance.
The reconfiguration schedule varies between designs: some achieve high performance but are impractical for large-scale deployment, while others offer more practical solutions but at the cost of moderate performance.

From a performance standpoint, traffic-oblivious networks, such as RotorNet~\cite{10.1145/3098822.3098838}, Sirius~\cite{10.1145/3387514.3406221}, and Opera~\cite{opera}, offer low reconfiguration overheads (in the range of nanoseconds) but sacrifice throughput due to their fixed and periodic switching schedules, which are independent of the underlying communication patterns. 
Only recently have the throughput bounds of traffic-oblivious networks been established~\cite{10.1145/3579312,10.1145/3519935.3520020}, showing that they are tightly bounded by $\frac{1}{2}$ \ie a  sustainable load of at most $50\%$ under worst-case traffic patterns (\eg ring communication), even with ideal routing and congestion control.
In contrast, traffic-aware networks such as Mordia~\cite{10.1145/2486001.2486007}, Helios~\cite{10.1145/1851182.1851223}, and ProjecToR~\cite{10.1145/2934872.2934911} are capable of achieving higher throughput because their switching schedules are optimized for the underlying communication patterns. Unfortunately, formal bounds on the achievable throughput of traffic-aware networks remains an open question in the literature.

From a practicality standpoint, periodic fixed-duration reconfigurations have emerged as a promising design choice for reconfigurable datacenter networks~\cite{Mellette:24,10.1145/3651890.3672273}. However, existing periodic networks are traffic-oblivious and require non-trivial additional support from hardware (NIC and switch) in terms of routing, packet reordering, congestion control and buffer architecture~\cite{10.1145/3651890.3672248,10.1145/3387514.3406221,10.1145/3098822.3098838}. This complexity hinders the large-scale deployment of these networks in practice. In contrast, traffic-aware networks typically only require single-hop (direct) routing and do not depend on in-network congestion control mechanisms. However, this comes at the cost of complex reconfiguration schedules with variable durations, and they often rely on an additional packet-switched network~\cite{10.1145/2486001.2486007,10.1145/2934872.2934911,10.1145/1851182.1851223}.

\medskip
In view of both performance and practicality, we explore a new direction in this paper: 
\emph{Can a high-throughput network be designed using periodic fixed-duration reconfigurations, without relying on multi-hop routing?}

\medskip
Figure~\ref{fig:intro} illustrates our perspective. Not only are existing designs limited in throughput, the worst-case throughput of \emph{any} traffic-oblivious network is bounded by $\frac{1}{2}$~\cite{10.1145/3519935.3520020,10.1145/3579312,10.1145/3491050}, making it infeasible to achieve higher throughput. However, it remains unexplored so far, whether and to what extent a traffic-aware approach to periodic reconfigurable networks can improve throughput. Intuitively, if the reconfiguration delay is negligible, a traffic-aware network can ideally achieve full-throughput. For instance, most prior works follow this intuition and use Birkhoff–von Neumann (BvN)~\cite{birkhoff1946three} decomposition technique to devise a circuit switching schedule that perfectly matches the underlying traffic pattern~\cite{10.1145/2486001.2486007,10.1145/2716281.2836126}. However, this not only results in a schedule with variable duration for each reconfiguration, it can also result in significantly low throughput due to reconfiguration overheads. Designing a traffic-aware network using only fixed-duration periodic reconfigurations, while surpassing the throughput limits of oblivious networks, requires new techniques and remains a challenging open problem.

We present \name, a first traffic-aware reconfigurable network design that breaks the throughput barrier of existing designs, using only fixed-duration periodic reconfigurations. \name not only achieves higher throughput but also greatly simplifies the protocol stack by eliminating multi-hop routing, congestion control and packet reordering. A recent work reports the following regarding single-hop routing in periodic circuit-switched network:

\medskip
\textit{``Notably,
none of this requires any modifications to the Linux application,
TCP, or the Linux networking stack.''~\cite{10.1145/3651890.3672273}}

\medskip
\name relies solely on direct communication. As a result, \name is more practical for deployment, within the available hardware capabilities \eg using Rotor switches~\cite{10.1145/3651890.3672273} and server-grade NICs~\cite{corundum}. We discuss \name's practicality as well as future research directions in more detail later in this~paper.

The key innovation behind \name is the use of a matrix rounding technique~\cite{bacharach1966matrix} to derive a switching schedule that matches the underlying traffic pattern. This is in contrast to BvN decomposition technique~\cite{birkhoff1946three} and greedy approximations~\cite{10.1145/2896377.2901479} that have been largely used in the literature for designing traffic-aware networks~\cite{10.1145/2486001.2486007,10.1145/2716281.2836126}.
Importantly, our rounding technique allows decomposing a traffic matrix (after rounding) into a set of permutation matrices that directly serve as the switching schedule with \emph{fixed-duration} for each circuit, in a periodic manner. Interestingly, the schedules produced by rounding technique are capable of serving a large portion of the demand but may leave certain residual demand. To this end, we introduce one extra cycle that is similar to that of existing designs, providing direct connections between all communicating nodes. Our rounding technique is simple and efficient to compute, making it practical for quick updates based on the changes in the traffic patterns. We establish the throughput bounds of \name, marking the first theoretical result on the achievable throughput of traffic-aware networks, while accounting for reconfiguration delays.

We evaluate using packet-level simulations and show that \name significantly improves throughput by up to $2.13$x and reduces the flow completion times by up to $99.54\%$ compared to existing approaches.

\begin{figure*}
\centering
\begin{minipage}{0.48\linewidth}
\centering
\includegraphics[width=1\linewidth]{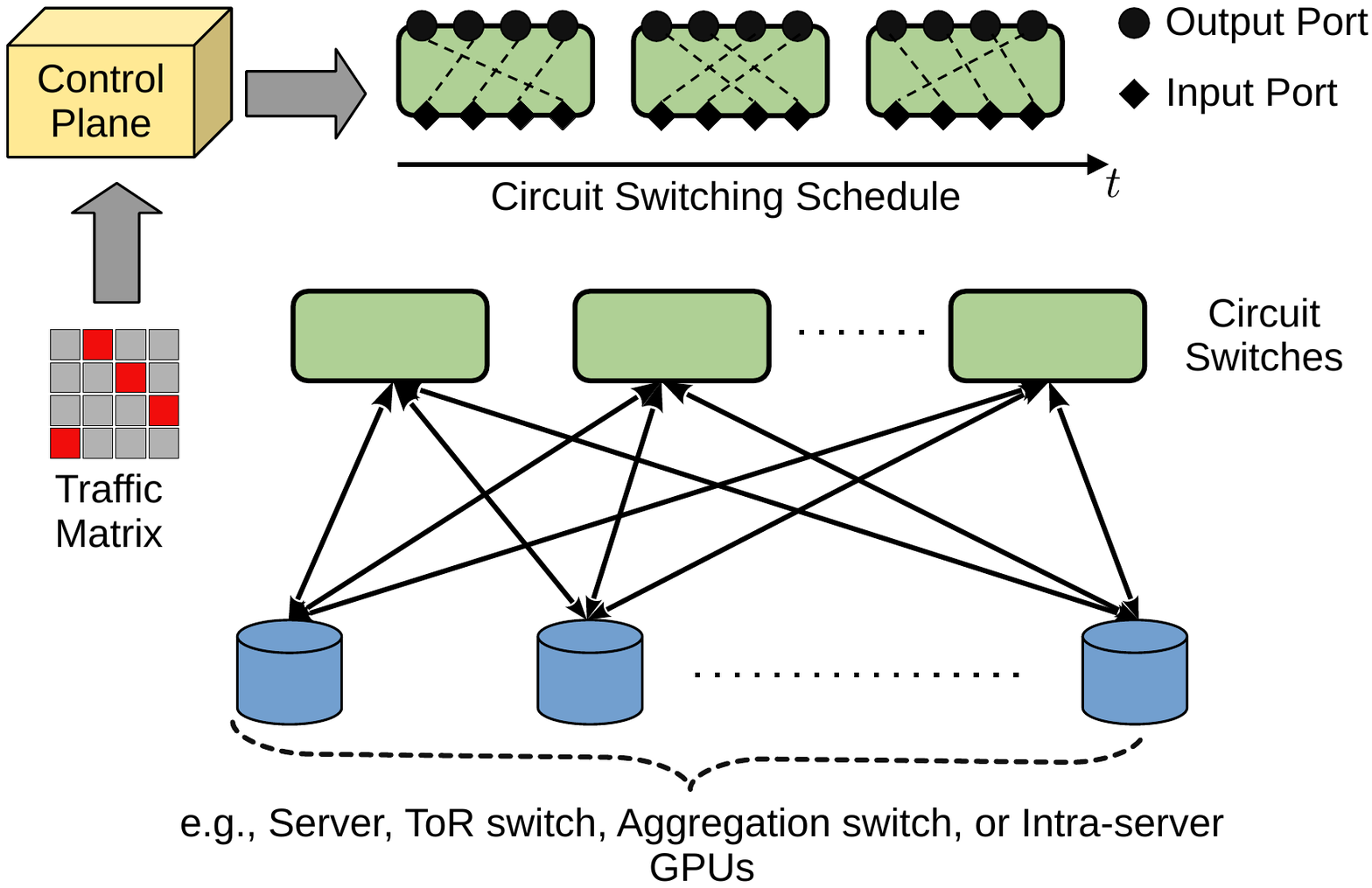}
\subcaption{The physical topology of a reconfigurable network consists of a set of nodes connected by optical circuit-switches arranged in a hierarchical Clos topology (\eg leaf-spine). The circuit-switches are time-synchronized and rapidly reconfigure their circuits providing direct links between pairs of nodes in a periodic manner. We assume a control plane that defines the switching schedule for each switch.}
\label{fig:topology}
\end{minipage}
\hfill
\begin{minipage}{0.48\linewidth}
\begin{subfigure}{1\linewidth}
\centering
\includegraphics[trim=0 8.6cm 0 0, clip,width=1\linewidth]{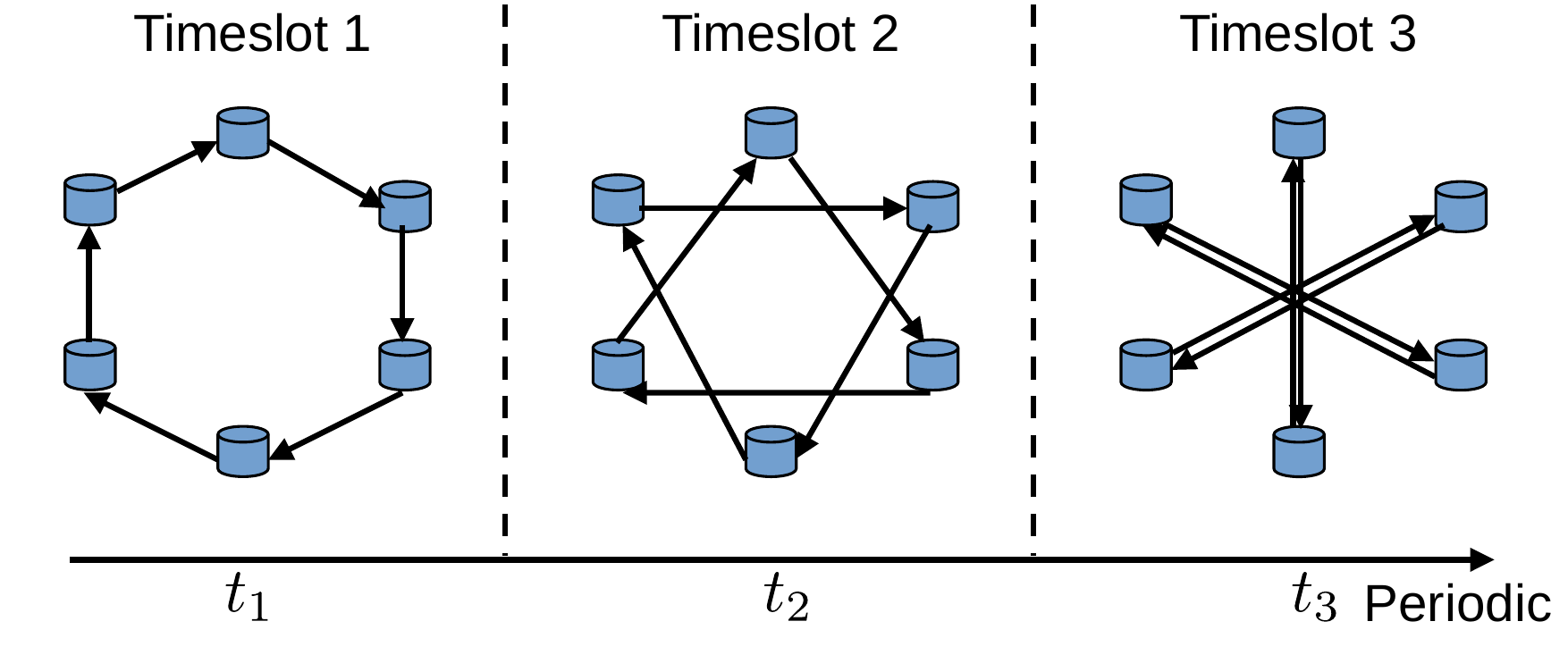}
\caption{The physical topology establishes a node-to-node network that evolves over time periodically, allocating bandwidth between pairs of nodes in each timeslot.\\}
\label{fig:temporal-graph}
\end{subfigure}
\begin{subfigure}{1\linewidth}
\centering
\includegraphics[trim=0 0 0 12cm, clip,width=1\linewidth]{figures/evolving-rdcn.pdf}
\caption{Examples of topologies that can be emulated over a \emph{period} using periodic circuit-switching}
\label{fig:emuated-graphs}
\end{subfigure}
\end{minipage}
\caption{The periodic circuit-switched network~(\ref{fig:topology}) rapidly reconfigures its circuits, forming a dynamic node-to-node topology that evolves over time~(\ref{fig:temporal-graph}). Over time, this topology emulates a specific static network~(\ref{fig:emuated-graphs}), for example, splitting bandwidth evenly across all pairs to create an all-to-all mesh topology (as in RotorNet~\cite{10.1145/3098822.3098838} and Sirius~\cite{10.1145/3387514.3406221}) or concentrating bandwidth between pairs with high demand to form a permutation or ring topology. The throughput of the network is heavily influenced by the choice of switching schedule, the emulated topology, and the underlying traffic matrix: \emph{an optimization opportunity}.}
\label{fig:rdcn}
\end{figure*}

Our main contributions in this work are:

\begin{itemize}[leftmargin=*,label=\small{\textcolor{myred}{$\blacksquare$}}]	
	\item A first separation result proving that traffic-aware reconfigurable datacenter networks are strictly superior to traffic-oblivious networks in terms of throughput.
    \item \name, an innovative, yet simple, traffic-aware network design based on periodic fixed-duration reconfigurations. \name achieves a throughput of at least $\frac{2}{3}$ (lower bound): a significant improvement over existing designs.
    \item A simplification of the requirements from network protocol stack. \name relies only on direct communication without multi-hop routing and does not need additional congestion control mechanisms to be deployed.
    \item Evaluations, highlighting the performance benefits of \name compared to traffic-oblivious counterparts. Our results show that \name significantly improves flow completion times for both short flows and long flows.
\end{itemize}

\emph{This work does not raise any ethical issues.}

\section{Motivation}
\label{sec:motivation}

In this section, we motivate our work by outlining the limitations of current periodic reconfigurable networks and the opportunities presented by adopting a traffic-aware approach within these systems.
Our primary focus in the rest of this paper is on \emph{periodic} reconfigurable networks, given the recent technological advancements showing their practicality and scalability~\cite{10.1145/3387514.3406221,10.1145/3651890.3672273}. In the following, we briefly describe the network architecture that we consider in this paper.

\medskip
\noindent
\textbf{Network model:}
Figure~\ref{fig:topology} illustrates the physical topology. A set of nodes are interconnected by optical circuit-switches such that at any time instance the network provides pair-wise direct connections across the nodes. For generality, we say ``nodes'' to refer to either servers, or ToR switches, or Aggregation switches, or intra-server components such as GPUs, that may be interconnected by an optical circuit-switched network.
The optical circuit-switches reconfigure according to a schedule in a synchronized and periodic manner. Specifically, the circuit-switches reconfigure at \emph{fixed-duration} intervals (timeslots) and each reconfiguration takes a specific amount of time (reconfiguration delay). This aligns with existing periodic circuit-switching technologies such as RotorNet~\cite{10.1145/3651890.3672273,10.1145/3098822.3098838,Mellette:24} (Sirius~\cite{10.1145/3387514.3406221}) with a reconfiguration delay of $7\mu s$ ($3.84ns$). As a result, the network topology evolves over time as shown in Figure~\ref{fig:temporal-graph}. In each timeslot, the degree of the topology is limited to the number of physical links. However, as the topology evolves over time, it can emulate a variety of topologies with high-degree, as shown in Figure~\ref{fig:emuated-graphs}. For example, the topology can emulate an all-to-all mesh, a ring, or a permutation topology, depending on the periodic circuit-switching schedule. 
So far, in the literature, periodic reconfigurable networks were only studied in a traffic-oblivious setup \ie the switching schedule is independent of the underlying traffic patterns, emulating an all-to-all mesh topology~\cite{10.1145/3098822.3098838,10.1145/3387514.3406221,opera} or a $d$-regular topology~\cite{10.1145/3651890.3672248,10.1145/3579312,10.1145/3519935.3520020} over time.

We first discuss our formal approach to optimizing the throughput of reconfigurable networks (\S\ref{sec:throughput}), followed by the drawbacks of existing traffic-oblivious designs (\S\ref{sec:drawbacks}), and later, we make a case for traffic-aware networks (\S\ref{sec:traffic-aware}).

\subsection{Throughput of Periodic Networks}
\label{sec:throughput}

Throughput offered by an interconnect is a crucial metric for assessing the sustainable load a network can handle, especially under highly concurrent communication patterns.
In order to quantify the throughput, we first formally define the communication pattern \ie the traffic matrix (Definition~\ref{def:demand-matrix}). The traffic matrix specifies the demand in bits per second between each pair of nodes \ie the total demand originating from a source towards a destination. Following prior work~\cite{10.1145/3452296.3472913}, we consider the hose model~\cite{10.1145/316188.316209} 
such that the total demand originating from (and destined to) each node is less than its corresponding capacity limits.

\begin{definition}[Traffic matrix]\label{def:demand-matrix}
    Given a set of nodes $\mathcal{N}$, each with $d$ outgoing and incoming links of capacity~$c$, a traffic matrix specifies the traffic rate between every pair of nodes in bits per second defined as $\mathcal{M}=\{ m_{u,v} \mid u\in \mathcal{N}, v\in \mathcal{N} \}$ where $m_{u,v}$ is the demand between the pair $u,v$. The traffic matrix is such that the total traffic originating at a source~$s$ is less than its outgoing capacity and the total demand terminating at a destination~$t$ is less than its incoming capacity \ie $\sum_{u\in \mathcal{N}} m_{s,u} \le c\cdot d$ and $\sum_{u\in \mathcal{N}} m_{u,t} \le c \cdot d$.
\end{definition}

For a given communication pattern and the corresponding traffic matrix (Definition~\ref{def:demand-matrix}), we define throughput as the maximum scaling factor such that there exists a feasible flow that can satisfy the scaled demand subject to flow conservation and capacity constraints. We denote flow by $F: P\mapsto \mathbb{R}^{+}$, a map from the set of all paths $P$ (static or temporal) to the set of non-negative real numbers. This mapping naturally ensures that the flow transmitted from a source eventually reaches the destination along a path $p\in P$. To obey capacity constraints, a feasible mapping is such that the sum of all flows traversing a link do not exceed the link capacity. We are now ready to define throughput formally.

\begin{definition}[Throughput]\label{def:throughput}
    Given a traffic matrix $\mathcal{M}$ and a reconfigurable network, throughput denoted by $\theta(\mathcal{M})$ is the highest scaling factor such that there exists a feasible flow for the scaled traffic matrix $\theta(\mathcal{M})\cdot \mathcal{M}$. Throughput~$\theta^*$ is the highest scaling factor for a worst-case traffic matrix \ie $\theta^* = \underset{\mathcal{M}\in \hat{M}}{\min} \theta(\mathcal{M})$, where $\hat{M}$ is the set of all demand matrices.
\end{definition}

Intuitively, throughput for a specific communication pattern captures the maximum sustainable load by the underlying topology. Based on Definition~\ref{def:throughput}, similar to prior works~\cite{10.1145/3452296.3472913,10.1145/3579312,10.1145/3491050,7877143}, throughput of a topology is the minimum throughput across the set of all saturated demand matrices \ie if a topology has throughput $\theta^*$, then it can achieve at least throughput~$\theta^*$ for \textit{any} traffic matrix and at most throughput~$\theta^*$ for a worst-case traffic matrix. 

Several variants of the throughput problem have been studied over the last decades, especially in the context of the maximum concurrent flow problem~\cite{10.1145/77600.77620}. However, in contrast to static networks, the fundamental challenge to study throughput in the context of reconfigurable networks is that the topology changes over time and can even be a function of the traffic matrix in the case of traffic-aware networks. 

In the specific case of periodic networks, a recent work establishes an equivalence between the throughput of any periodic network and a corresponding emulated topology (as static graph), allowing the study of throughput in periodic networks using existing techniques~\cite{10.1145/3579312}. We present a formal definition of emulated graph in Appendix~\ref{app:throughput}. Essentially, the emulated graph is a time-collapsed view over an entire period of the periodic network as shown in Figure~\ref{fig:emuated-graphs}. 

For example, in RotorNet, the length of the period is $\Gamma=\frac{n}{d}$ timeslots. If a link appears once between every pair of nodes over a period, then the emulated graph has one link between every pair of nodes (complete graph), where each link has a capacity of $c\cdot \frac{1}{\Gamma} = c\cdot \frac{d}{n}$. As we will discuss later, emulating a complete graph results in a drop in throughput by a factor~$2$ under any permutation (\eg ring) communication patterns. Intuitively, in order to maximize throughput, the emulated graph must provide high bandwidth between specific pairs of nodes with high-demand \ie by emulating a topology that closely matches the underlying communication pattern.

\medskip
\noindent{\textcolor{myred}{$\blacksquare$ \textbf{\textit{Takeaway.}}} \textit{
Optimizing periodic networks entails finding the best static graph that can be emulated over a period, that provides the highest throughput for an (any) underlying communication pattern.
}}

\begin{figure*}
\centering
\begin{minipage}{0.20\linewidth}
\centering
\includegraphics[trim= 0 9cm 24cm 0, clip, width=0.75\linewidth]{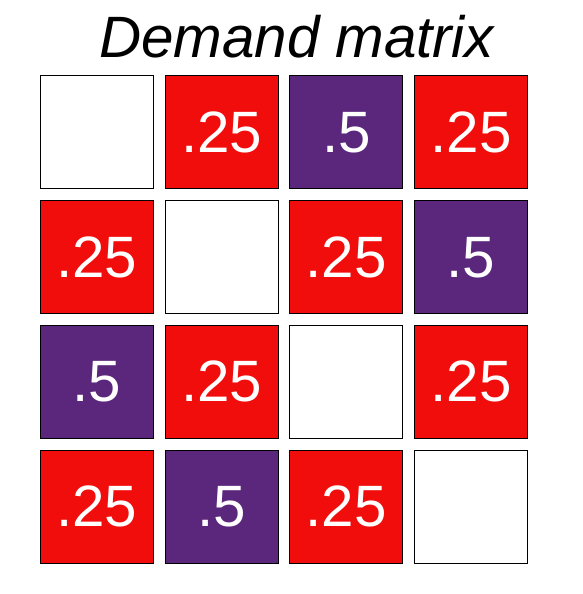}
\subcaption{Example traffic matrix for a topology with $4$ nodes, each with $1$ physical link. }
\label{fig:example-matrix}
\end{minipage}\hfill
\begin{minipage}{0.25\linewidth}
\centering
\includegraphics[trim= 23cm 0cm 0 0, clip, width=0.63\linewidth]{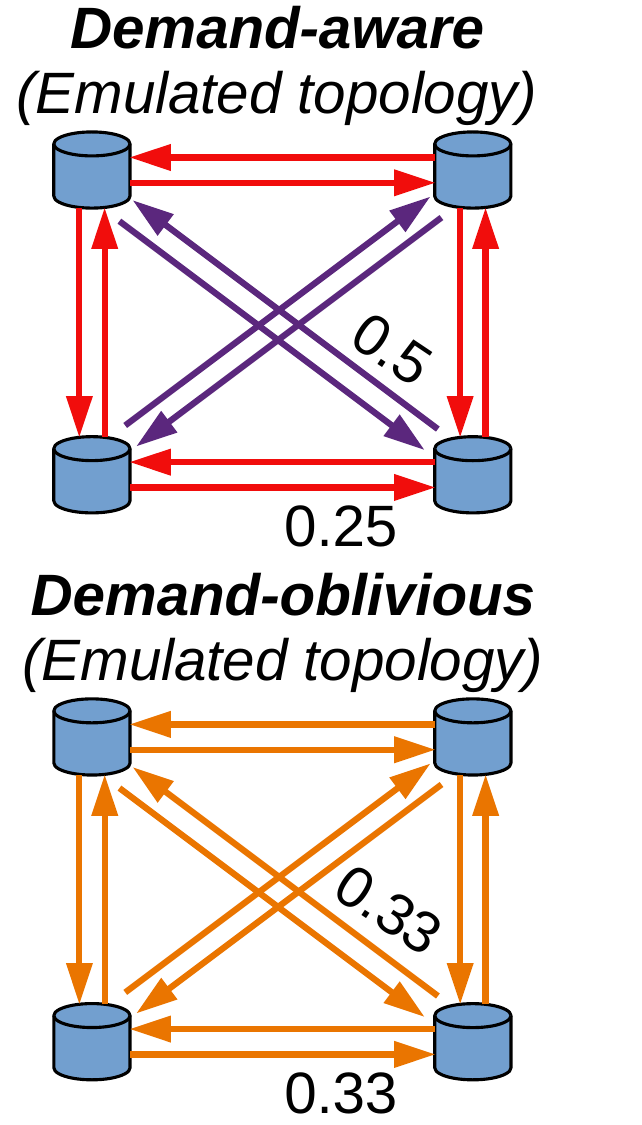}
\subcaption{Target emulated topology for traffic-aware vs oblivious periodic networks.}
\label{fig:target-emulated}
\end{minipage}\hfill
\begin{minipage}{0.50\linewidth}
\centering
\includegraphics[width=1\linewidth]{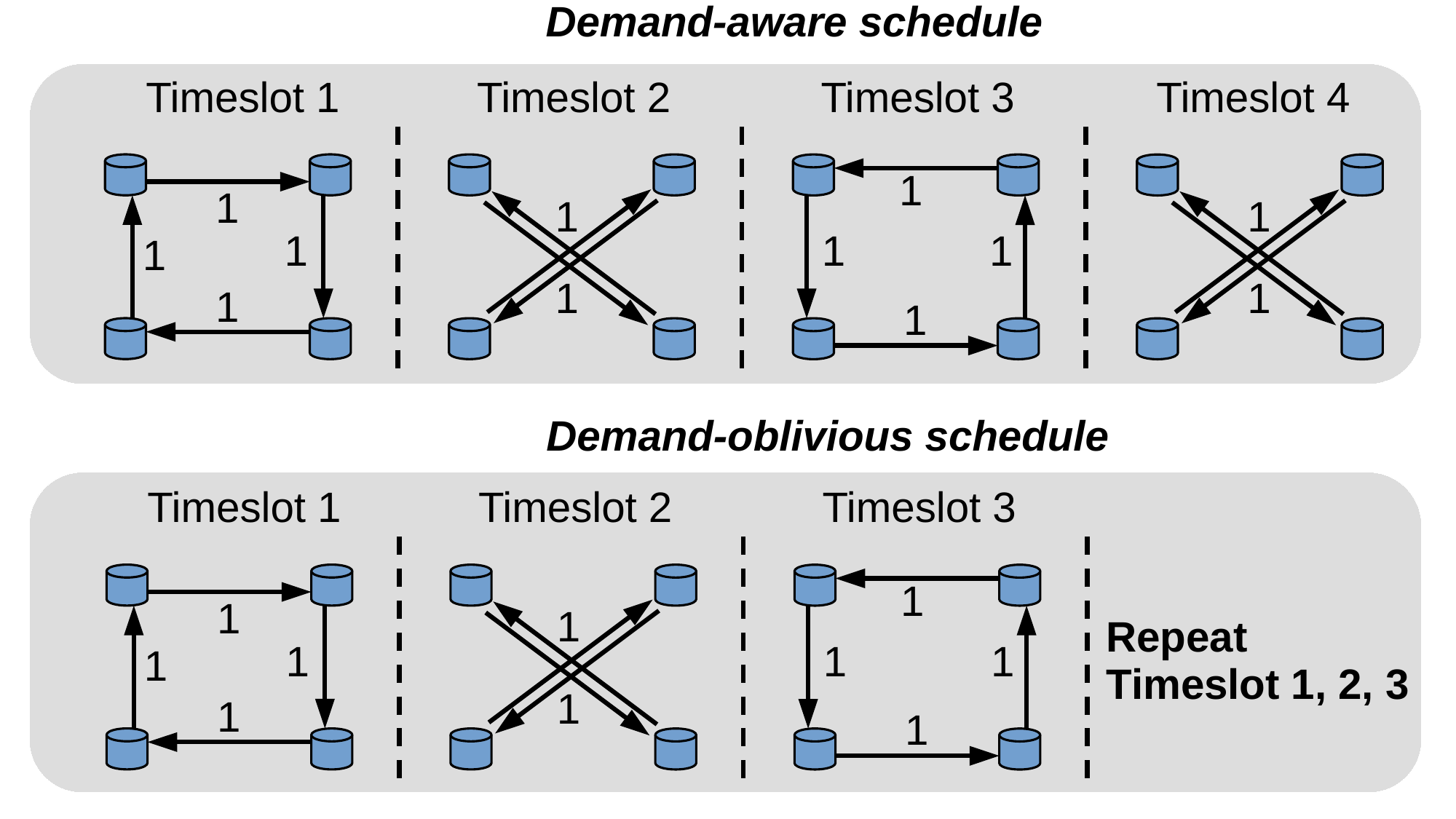}
\subcaption{Derived periodic schedule for a traffic-aware vs oblivious periodic network.}
\label{fig:example-schedules}
\end{minipage}\hfill
\caption{Interpreting the traffic matrix (Figure~\ref{fig:example-matrix}) as a target emulated graph (Figure~\ref{fig:target-emulated}) allows deriving a periodic schedule (Figure~\ref{fig:example-schedules}) that can achieve full throughput for traffic-aware periodic networks, even with single-hop routing. In contrast, the schedule for an oblivious periodic network often does not match the underlying traffic matrix, necessitating multi-hop routing and resulting in reduced throughput.}
\label{fig:example}
\end{figure*}

\subsection{Drawbacks of ORNs}
\label{sec:drawbacks}
Oblivious periodic circuit-switching has an obvious advantage of simplicity without any control plane involvement for optimizing the topology. However, this leads to certain drawbacks that we elaborate in the following.

\medskip
\noindent
\textbf{Factor of $2$ reduction in throughput:} Recent works have established a tight bound of $\frac{1}{2}$ on the throughput of periodic networks~\cite{10.1145/3519935.3520020,10.1145/3579312,10.1145/3491050}. We emphasize that this reduction in throughput is not solely attributed to the use of valiant load-balancing; rather, the network fundamentally cannot sustain beyond $50\%$ load, even with an ideal routing scheme, under a worst-case communication pattern such as a ring. Specifically, in ring communication, only specific node pairs exchange data, but emulating a complete graph only offers a capacity of $c \cdot \frac{d}{n}$ (as discussed above), falling short of the ideal bandwidth provisioning of $c \cdot d$ (leveraging all available links) for these pairs. Even an optimal routing scheme is thus forced to route traffic over $2$-hop paths, in order to fully utilize the network capacity, leading to a throughput of $\frac{1}{2}$.

\medskip
\noindent
\textbf{Multi-hop routing:} The use of indirect paths is a necessity to improve utilization in oblivious periodic networks. However, since the network is evolving over time, this implies that intermediate nodes need to buffer (hold) traffic until the next link along the path becomes available. This results in additional latency and buffer requirements. In addition, the use of multi-hop routing requires the use of in-network congestion control mechanisms to prevent packet loss due to buffer overflows~\cite{10.1145/3651890.3672248}. This further complicates the network stack and can lead to performance degradation under high load. In fact, recent works even suggest sacrificing throughput further in order to satisfy buffer constraints~\cite{10.1145/3579312}.

\medskip
\noindent
\textbf{Packet reordering:} Not only do existing oblivious networks require multi-hop routing, but they also require multi-path transmission in order to fully utilize the network. This results in packets arriving out-of-order at the destination, necessitating additional support such as reorder buffers~\cite{10.1145/3651890.3672273}. This can be particularly challenging in the context of RDMA since RoCE implementations typically react to packet reordering at the receiver with NACK that triggers retransmissions immediately at the sender.

\medskip
\noindent{\textcolor{myred}{$\blacksquare$ \textbf{\textit{Takeaway.}}} \textit{
While oblivious periodic networks simplify switching schedule selection, they rely on complex protocols such as multi-hop routing, multi-path transmission, and reorder buffers, which can significantly reduce the maximum achievable throughput.
}}

\subsection{A Case for traffic-aware Networks}
\label{sec:traffic-aware}

In contrast to oblivious reconfigurable networks, traffic-aware networks can potentially achieve higher throughput by optimizing the network topology for the underlying communication pattern. traffic-aware networks have been studied in the literature~\cite{10.1145/2934872.2934911,10.1145/1851182.1851223,10.1145/2486001.2486007} and have been empirically shown to achieve better performance compared to oblivious networks. 
In fact, traffic-aware networks can satisfy any demand within the hose model, if the reconfiguration delay is hypothetically near-zero or negligible. This result has been intuitively known in the literature~\cite{10.1145/2486001.2486007,10.1145/3409964.3461786}, which relies on Birkhoff-von-Neumann (BvN) matrix decomposition technique~\cite{birkhoff1946three}. For completeness, we formally state it here with a proof in Appendix~\ref{app:throughput}.

\begin{restatable}[Ideal throughput of traffic-aware network]{theorem}{idealThroughputTheorem}\label{th:throughput-ideal}
The throughput of an ideal traffic-aware reconfigurable network is $1$ \ie full-throughput for any traffic matrix if the reconfiguration delay is negligible. 
\end{restatable}

The core intuition behind BvN-based traffic-aware network design is to allocate bandwidth between node-pairs as a convex combination (over time) of permutations derived from BvN matrix decomposition. However, this approach suffers from two key limitations: \first, the durations between reconfigurations in the switching schedule are variable, and \second, the time between reconfigurations can be shorter than the reconfiguration delay, leading to a significant drop in throughput. Consequently, BvN-based designs are incompatible with current technologies that rely on fixed-duration periodic reconfigurations.

Interestingly, it is possible to design periodic reconfigurable networks that achieve full throughput for specific types of communication patterns. 
For instance, consider a ring traffic matrix. Even with fixed-duration periodic reconfigurations, a simple switching schedule with a period of one timeslot --- where the direct links between communicating nodes in the ring are maintained --- can achieve full throughput. In contrast, oblivious periodic networks can only achieve a throughput of~$\frac{1}{2}$ for the same communication pattern. Similarly, if the traffic matrix consists of integer multiples of link capacity, it is straightforward to design a switching schedule that achieves almost full throughput. Simply establishing direct links between communicating nodes is sufficient to achieve full throughput for such traffic matrices. 

Our key insight is that when the periodic schedule is constrained to $\Gamma$ timeslots, any traffic matrix $\mathcal{M}$ with non-zero demand represented as an integer multiple of $\frac{c}{\Gamma}$ can be satisfied with full throughput. For such communication patterns, a feasible periodic schedule with fixed-duration reconfigurations always exists and achieves full throughput. We formally prove this result in Appendix~\ref{app:throughput}. Intuitively, the traffic matrix can be visualized as an edge-weighted graph, where the weights represent the demand between node pairs. This graph corresponds to the target emulated topology for the periodic schedule, with weights now representing link capacities. A periodic schedule can then be derived to allocate bandwidth between source-destination pairs according to the traffic matrix, ensuring full throughput.

\begin{restatable}[Throughput under integer traffic matrices]{theorem}{intergerMatrixTheorem}\label{th:throughput-integer}
There exists a periodic reconfigurable network with a period of $\Gamma$ timeslots with $\Delta_r$ fraction of time spent in reconfiguration, that can achieve nearly full throughput of $1-\Delta_r$ using only single-hop routing for traffic matrices where any non-zero demand is an integer multiple of $c\cdot \frac{1}{\Gamma}$, where $c$ is the link capacity in the physical topology.
\end{restatable}

Figure~\ref{fig:example} illustrates a traffic matrix for which a traffic-aware periodic schedule can be trivially derived. The traffic matrix consists of integer multiples of $\frac{1}{4}$, and the corresponding switching schedule, with a length of 4 timeslots, is depicted in Figure~\ref{fig:example}. This schedule enables the network to fully satisfy the traffic matrix within each period, achieving full throughput. In contrast, the schedule of an oblivious network contains only 3 timeslots, omitting the additional $4^{th}$ timeslot used in the traffic-aware schedule. The oblivious schedule distributes uniform bandwidth across all node-pairs but mismatches the underlying traffic matrix, as demonstrated in Figure~\ref{fig:example}. Notably, the inclusion of just a few extra timeslots in a traffic-aware manner can significantly enhance the network's throughput.

\medskip
\noindent
\textbf{Single-hop routing:} 
Notice that the traffic-aware schedule in Figure~\ref{fig:example} achieves high throughput even with single-hop routing. This is because the emulated topology provides capacity between each node pair that precisely matches the traffic matrix specifications. Generally, any schedule derived from Theorem~\ref{th:throughput-integer} maintains this single-hop routing advantage, simplifying the protocol stack by eliminating the need for multi-hop routing, congestion control, and packet reordering.

However, a caveat is that increasing the period $\Gamma$ can lead to an excessively long periodic schedule to achieve full throughput for any traffic matrix, as indicated by Theorem~\ref{th:throughput-integer}. This can result in unacceptable delays. Theorem~\ref{th:throughput-integer} suggests potential throughput gains by deriving periodic schedules tailored to the underlying communication patterns. The challenge remains in deriving \emph{short} schedules while still achieving high throughput.

\medskip
\noindent{\textcolor{myred}{$\blacksquare$ \textbf{\textit{Takeaway.}}} \textit{
Specific communication patterns showcase the substantial throughput gains that traffic-aware periodic networks can achieve over oblivious designs. The main challenge --- and opportunity --- lies in deriving compact schedules that can achieve high throughput for any communication pattern.
}}

\section{traffic-aware Periodic Network}
\label{sec:design}

Based on our observations in \S\ref{sec:motivation}, we seek to design a \emph{simple} traffic-aware periodic network, within the practical capabilities of existing optical circuit-switching technologies~\cite{10.1145/3651890.3672273,Mellette:24,10.1145/3387514.3406221} and the end-host networking stack.
Our goal is to achieve high throughput for any traffic matrix using only single-hop routing \ie direct communication. 
We first present our network design (\S\ref{sec:vermilion}), followed by the throughput guarantees of our design (\S\ref{sec:properties}) and its practicality (\S\ref{sec:practicality}).

\subsection{Vermilion}
\label{sec:vermilion}

We present \name, a first traffic-aware periodic network design that can probably achieve high throughput compared to existing oblivious designs. We walk through each component of \name in the following.

\medskip
\noindent
\textbf{Physical topology:} Our network model remains the same as described in \S\ref{sec:motivation}, with a set of nodes interconnected by optical circuit-switches in a hierarchical CLOS topology.
The circuit-switches are synchronized in time and reconfigure at fixed-duration intervals, forming a dynamic node-to-node topology that evolves over time. Each node in the topology has $\hat{d}$ physical links that connect to the optical interconnect, hence at any time instance, each node can connect to at most $\hat{d}$ other nodes. The physical links have a capacity of $c$.

\begin{algorithm}[t]
	\SetKwFunction{emulatedTopology}{\textbf{\textsc{\textcolor{myred}{emulatedTopology}}}}
	\SetKwFunction{generateSchedule}{\textbf{\textsc{\textcolor{myred}{generateSchedule}}}}

	\SetKwProg{Fn}{function}{:}{}
	\SetKwProg{Proc}{procedure}{:}{}
	\SetKwInOut{KwIn}{Input}
	\SetKwInOut{KwOut}{Output}

	\KwIn{\ Traffic matrix $\mathcal{M}$, number of nodes $n$, \\ \ \ 
	degree $\hat{d}$, link capacity $c$, parameter $k$}

	\Proc{\generateSchedule{$\mathcal{M}$}}{

		$G=$ emulatedTopology($\mathcal{M}$)

		\Comment{\textcolor{gray}{\textit{Sequence of matchings for the periodic schedule}}}

		\For{$i=1$ \KwTo $k\cdot n$}{

			remove one perfect matching $M_i$ in $G$

			add $M_i$ to schedule
		}
		
		\KwRet $M$
	}

	\Proc{\emulatedTopology{$\mathcal{M}$}}{

		\Comment{\textcolor{gray}{\textit{Initialize a multigraph}}}
		
		$G=(V, E)$, $V = \{1, ..., n\}$, $E: V\times V \rightarrow \mathbb{N}$

		Normalize the traffic matrix $\mathcal{M}$

		$\mathcal{M} \leftarrow (k-1)\cdot n \cdot \mathcal{M}$
		
		$\mathcal{R} =$ Round($\mathcal{M}$) \Comment{\textcolor{gray}{\textit{Matrix rounding}}}
		
		\For{each node pair $(u,v)$}{
			
			\Comment{\textcolor{gray}{\textit{Allocate bandwidth for bulk demand}}}
			
			$E((u,v)) \leftarrow \mathcal{R}(u,v)$ \Comment{\textcolor{gray}{\textit{\# Edges between $u,v$}}}
			
			\Comment{\textcolor{gray}{\textit{Allocate bandwidth for residual demand}}}

			$E((u,v)) \leftarrow E((u,v)) + 1$
		}

		\Comment{\textcolor{gray}{\textit{Ensure that the final graph is regular}}}

		$x^{in} \leftarrow k\cdot n - x^{in}$ \Comment{\textcolor{gray}{\textit{Remaining in-degree}}}
		
		$x^{out} \leftarrow k\cdot n - x^{out}$ \Comment{\textcolor{gray}{\textit{Remaining out-degree}}}

		$G^{\prime} = (V, E^{\prime})=$ ConfigurationModel($x^{in}, x^{out}$)
		
		$E \leftarrow E \uplus E^{\prime}$

		\KwRet $G$

	}

	\caption{\name}
	\label{alg:vermilion}
\end{algorithm}

\medskip
\noindent
\textbf{Parameters:}
\name has one parameter $k$, that controls the degree of the target emulated topology.  A higher $k$ leads to higher throughput but also increases the schedule length. 
The resulting schedule can achieve at least $\frac{k-1}{k}$ throughput for a given traffic matrix $\mathcal{M}$. For example, even with $k=3$, \name can achieve a throughput of $\frac{2}{3}$, breaking the throughput bounds of oblivious periodic networks. We discuss the choice of $k$ in \S\ref{sec:properties}.

\begin{figure*}
\centering
\begin{subfigure}{0.325\linewidth}
\centering
\includegraphics[width=0.8\linewidth]{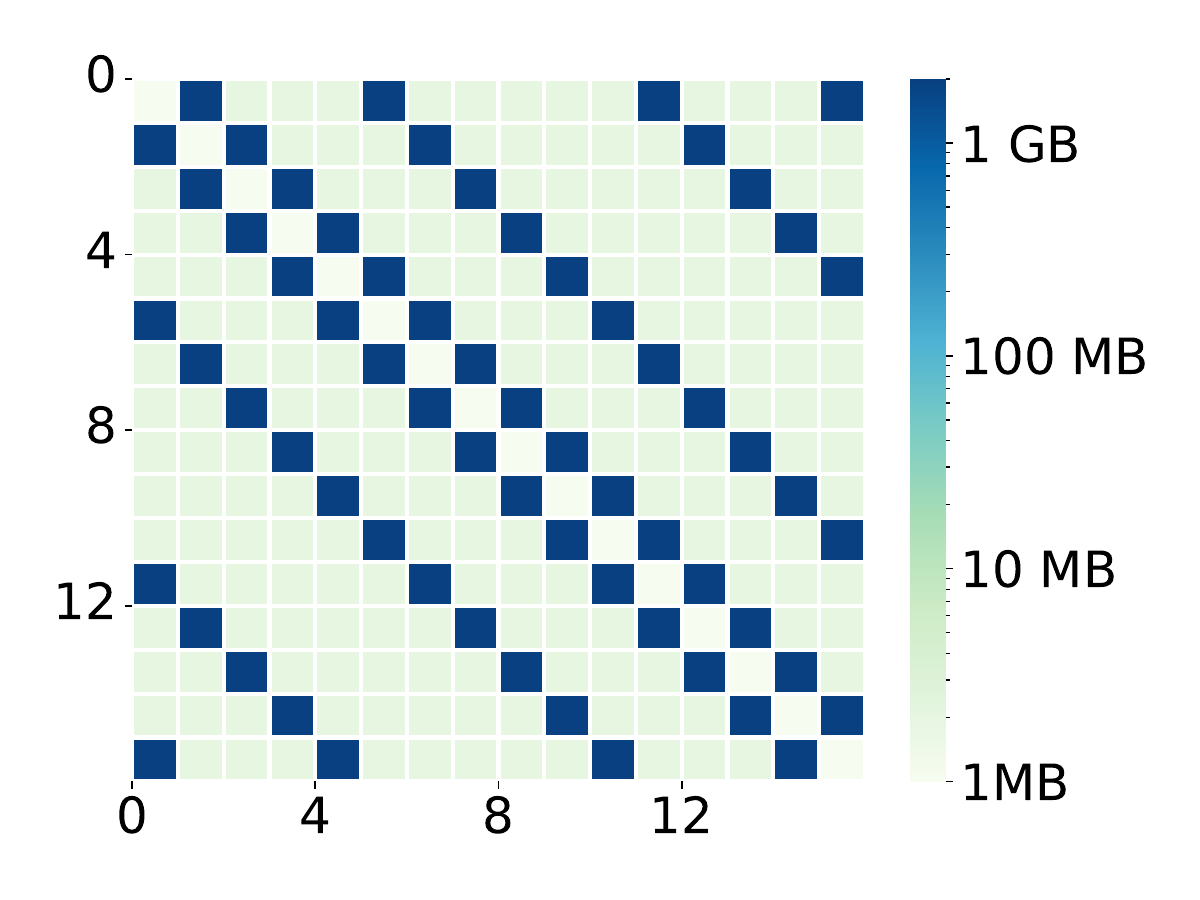}
\caption{DLRM data parallelism traffic matrix}
\label{fig:dlrm-heatmap}
\end{subfigure}\hfill
\begin{subfigure}{0.333\linewidth}
\centering
\includegraphics[width=0.8\linewidth]{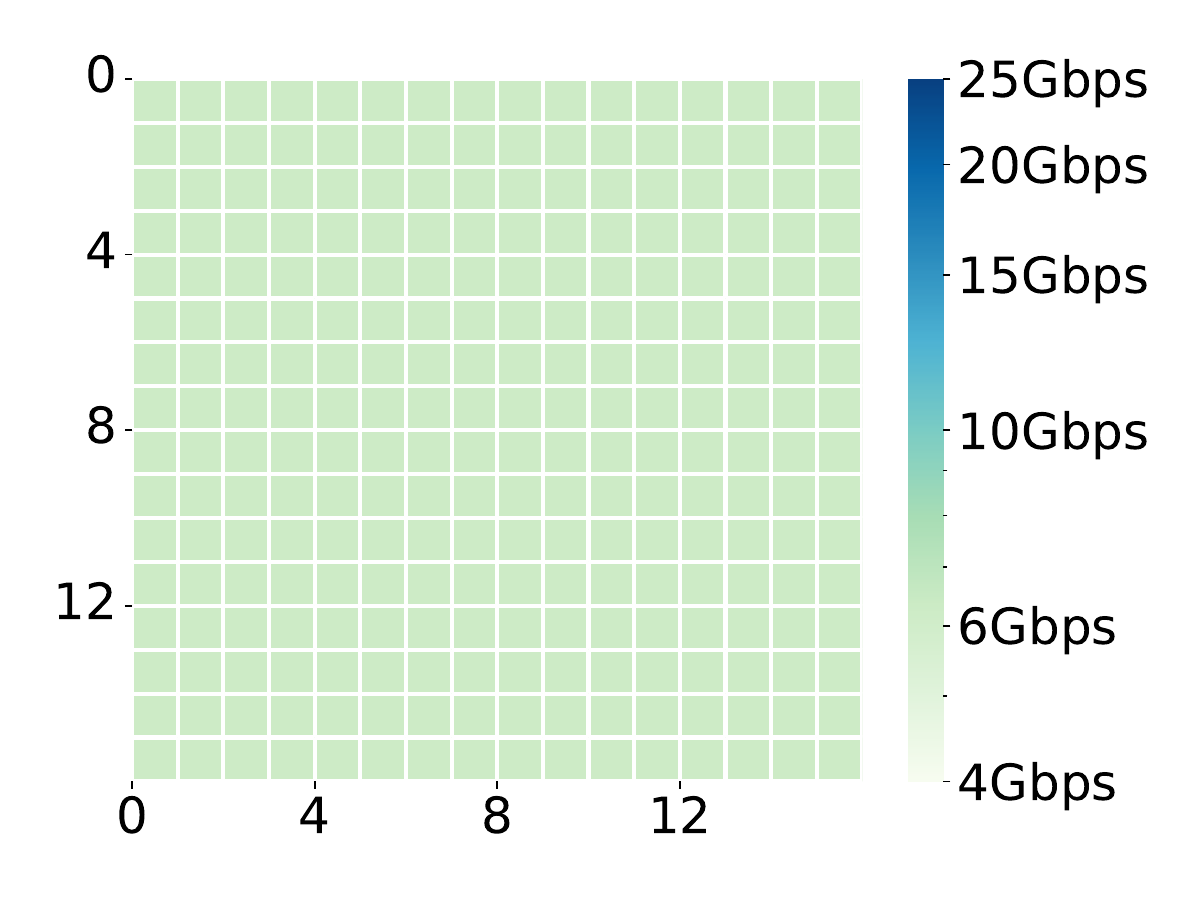}
\caption{Capacity provided by Oblivious network}
\label{fig:example-emulated-oblivious}
\end{subfigure}
\begin{subfigure}{0.325\linewidth}
\centering
\includegraphics[width=0.8\linewidth]{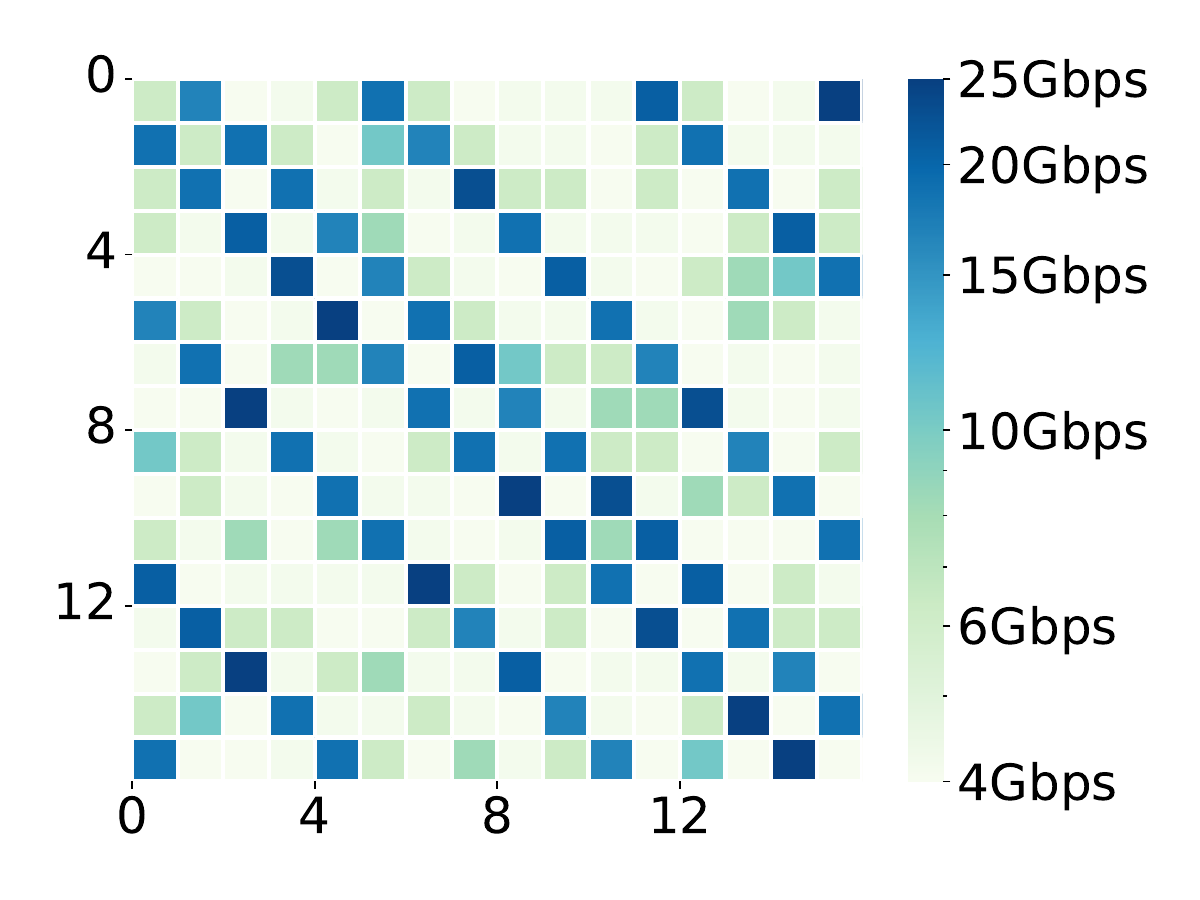}
\caption{Capacity provided by \namebold}
\label{fig:example-emulated-vermilion}
\end{subfigure}
\begin{subfigure}{0.19\linewidth}
\centering
\includegraphics[width=1\linewidth]{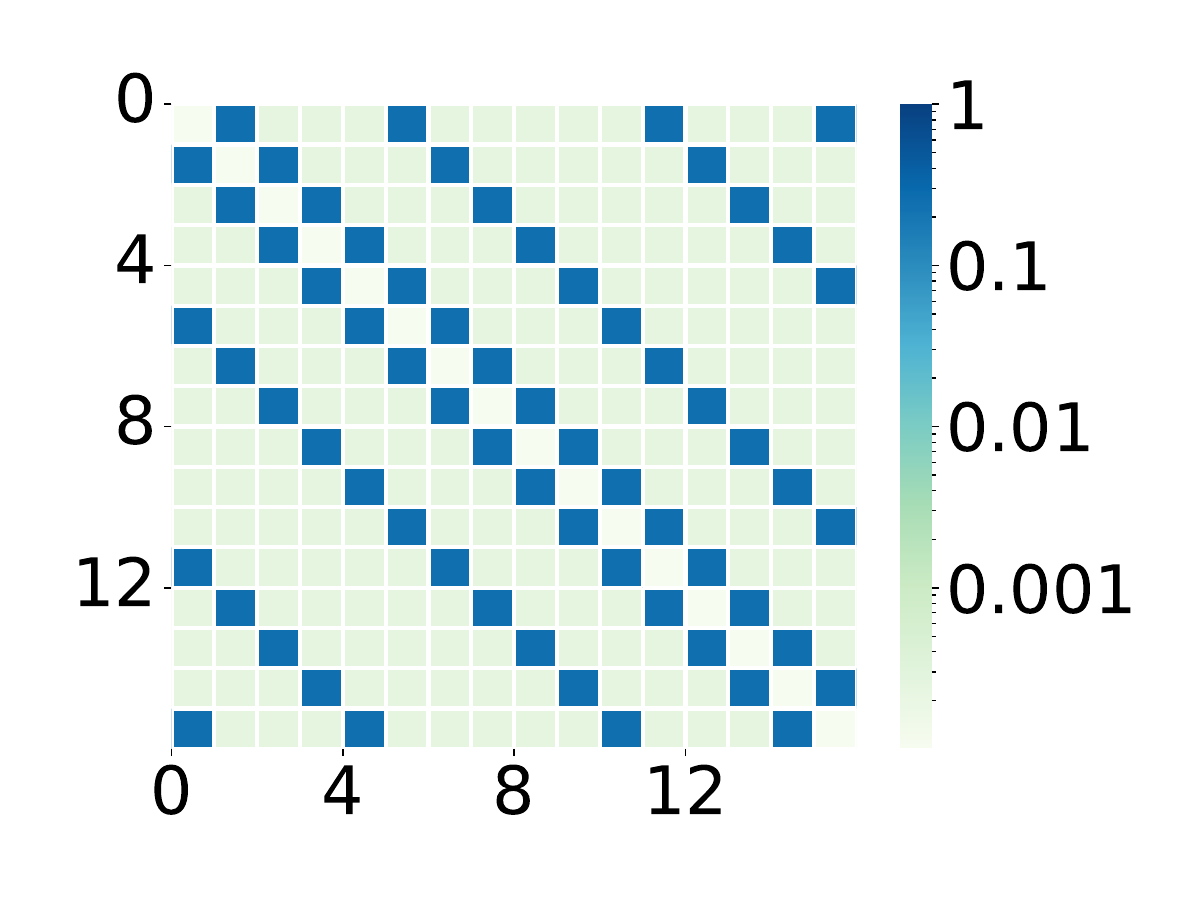}
\caption{Normalized matrix}
\label{fig:example-normalized}
\end{subfigure}\hfill
\begin{subfigure}{0.19\linewidth}
\centering
\includegraphics[width=1\linewidth]{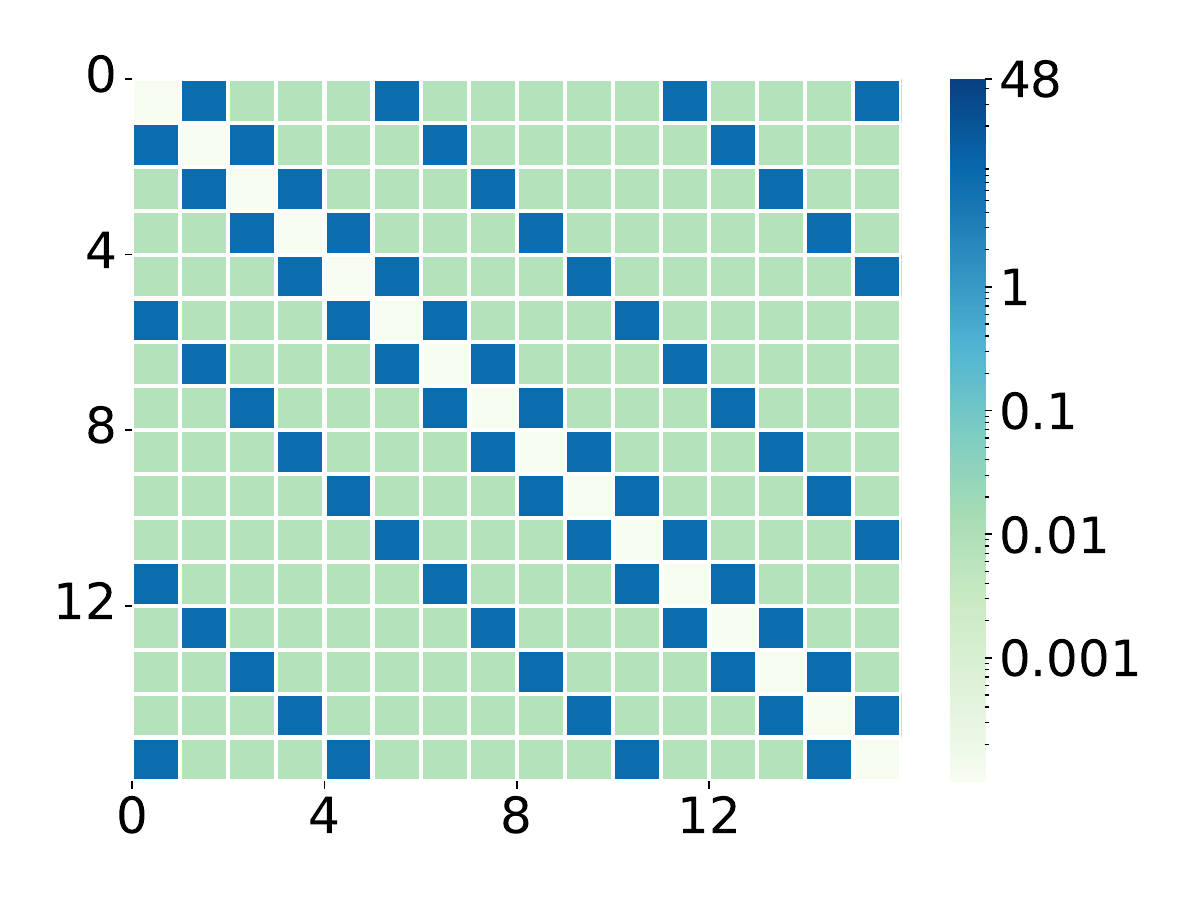}
\caption{Scaled by $(k-1)\cdot n$}
\label{fig:example-upscaled}
\end{subfigure}\hfill
\begin{subfigure}{0.19\linewidth}
\centering
\includegraphics[width=1\linewidth]{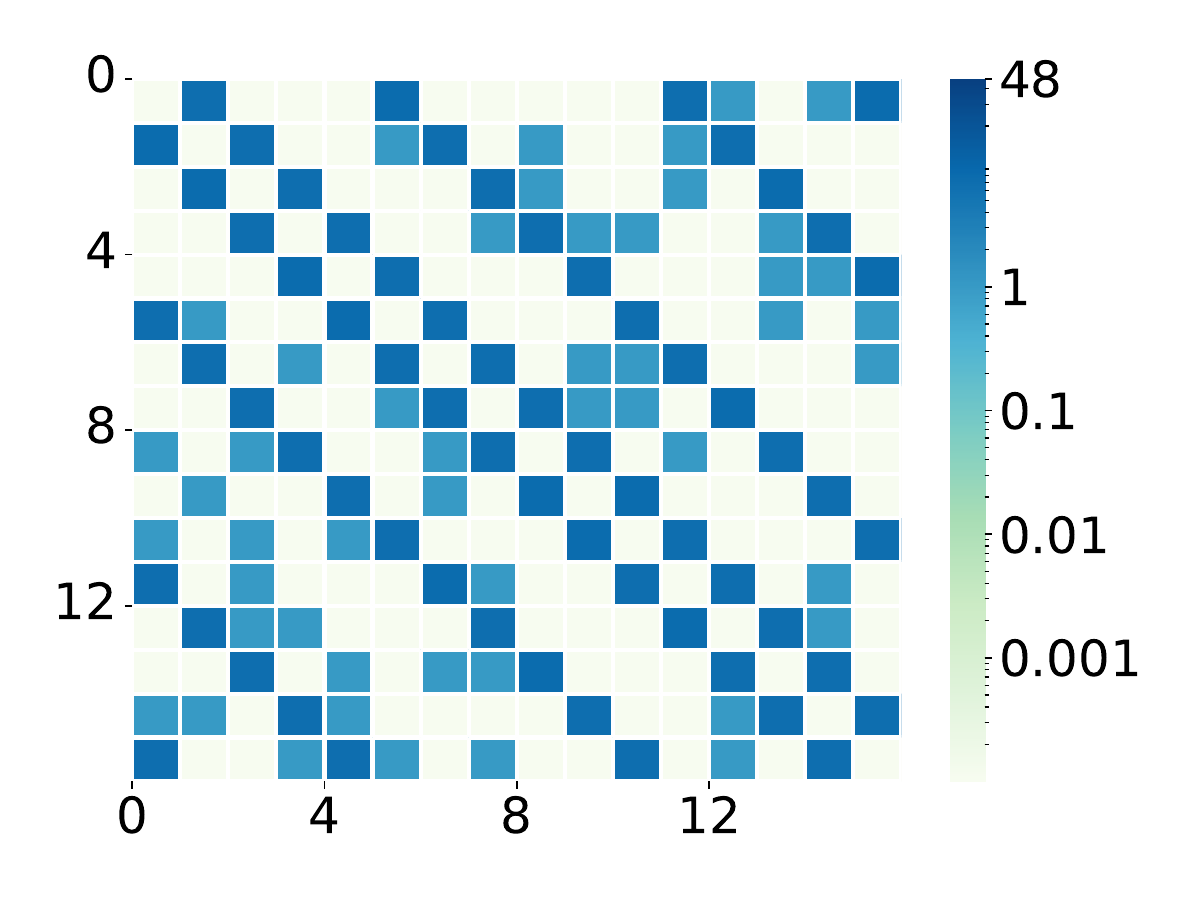}
\caption{Rounded matrix}
\label{fig:example-rounded}
\end{subfigure}\hfill
\begin{subfigure}{0.19\linewidth}
\centering
\includegraphics[width=1\linewidth]{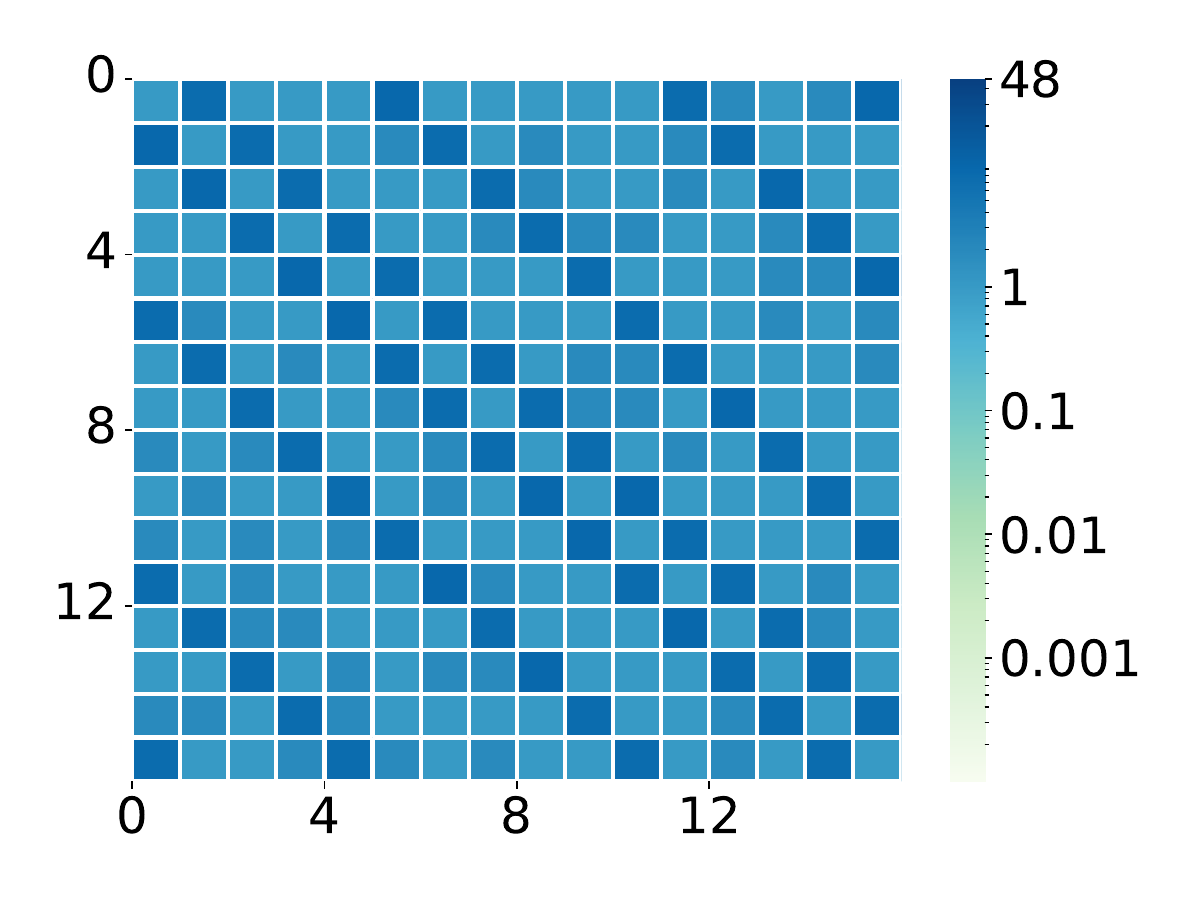}
\caption{$\uplus$ complete graph}
\label{fig:example-completeaug}
\end{subfigure}\hfill
\begin{subfigure}{0.2\linewidth}
\centering
\includegraphics[width=1\linewidth]{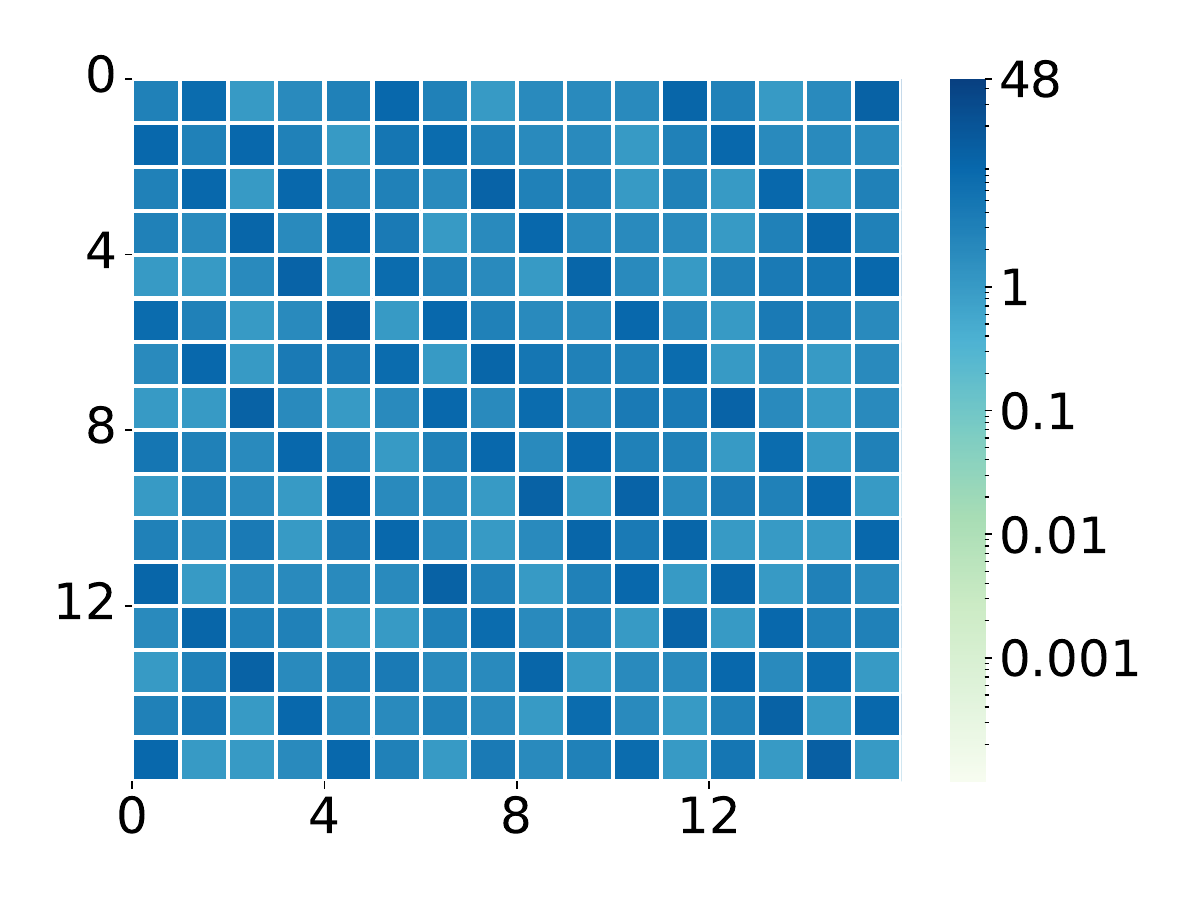}
\caption{$\uplus$ configuration model}
\label{fig:example-configurationaug}
\end{subfigure}
\caption{An example workflow of \name with~(\ref{fig:dlrm-heatmap}) DLRM data parallel traffic matrix. An oblivious periodic network~(\ref{fig:example-emulated-oblivious}) provides uniform capacity between all node pairs irrespective of the traffic matrix. \name~(\ref{fig:example-emulated-vermilion}) provides capacity between node pairs based on the underlying traffic matrix. \name first~(\ref{fig:example-normalized}) normalizes the given traffic matrix, upscales it by $(k-1)\cdot n$~(\ref{fig:example-upscaled}), rounds it~(\ref{fig:example-rounded}), augments it with a complete graph~(\ref{fig:example-completeaug}) for any-to-any connectivity, and finally augments it with additional links to ensure regularity~(\ref{fig:example-configurationaug}). The resulting matrix (a multigraph) is the target emulated topology~(\ref{fig:example-emulated-vermilion}), which is then decomposed into a sequence of matchings for the periodic schedule.}
\label{fig:dlrm-example}
\end{figure*}

\medskip
\noindent
\textbf{Periodic schedule:}
The key innovation in \name is its traffic-aware periodic schedule.
Algorithm~\ref{alg:vermilion} outlines the steps to derive the periodic schedule. Figure~\ref{fig:dlrm-example} shows an example workflow of \name for $k=3$. 
Given a traffic matrix~$\mathcal{M}$, we first generate an emulated topology that can achieve high throughput for the given traffic matrix. Our emulated topology is always regular and allows us to then decompose it into a periodic schedule.
We construct the emulated topology as follows:

\begin{itemize}[label=\small{\textcolor{myred}{$\blacksquare$}}]
	\item \textbf{Matrix rounding:} 
	We first normalize the traffic matrix such that the maximum sum of any row and column is at most $1$. We then upscale the traffic matrix by $(k-1)\cdot n$, where $k$ is a parameter to \name and $n$ is the number of nodes. 
	We round entries of the scaled matrix such that the sum of each row and column remains the same, a technique known as matrix rounding~\cite{bacharach1966matrix}. 

  \smallskip
	\item \textbf{traffic-aware multigraph:} Based on the rounded matrix, we construct a multigraph by adding edges between each node pair based on the rounded matrix. For instance if the rounded matrix specifies $2$ between $s$-$t$ node pair, then we add $2$ edges between $s$ and $t$. This ensures that majority of the traffic matrix is served efficiently in a traffic-aware manner.

	\smallskip
  \item \textbf{traffic-oblivious residual graph:} Matrix rounding may not exactly match the original traffic matrix and can leave some residual demand. To address this, we ensure that any residual demand can be routed by adding one additional edge between each node pair in the multigraph. This step guarantees any-to-any connectivity and is traffic-oblivious, meaning it is independent of the specific traffic matrix.

	\smallskip
  \item \textbf{Augmenting the regularity of the graph:}
	At this point, say each node has in-degree $x^{in}_i$ and out-degree $x^{out}_i$, we then take the degree sequences $\langle k\cdot n - x^{in}_1, ..., k\cdot n - x^{in}_n \rangle$ and $\langle k\cdot n - x^{out}_1, ..., k\cdot n - x^{out}_n \rangle$, and add additional links to our graph based on configuration model using the above degree sequence. 
\end{itemize}

The resulting graph is our target emulated topology. 
The above construction always leads to a multigraph with degree $k\cdot n$, a directed regular graph, that can be decomposed into $k\cdot n$ number of perfect matchings. These perfect matchings are then executed in round-robin periodically, using optical circuit switching. 

\medskip
\noindent
\textbf{Routing:} \name relies solely on single-hop routing, as the topology provides sufficient direct links between communicating nodes over time, ensuring high throughput even with direct communication.

\medskip
\noindent
\textbf{Congestion control:} \name does not explicitly require any congestion control algorithm in the network. If the network is all-optical \ie when servers connect directly to circuit switches, then \name does not require even end-host congestion control (except for reliability) since every packet is transmitted directly to the destination. The direct communication paths ensure that the network is not congested. 

\medskip
\noindent
\textbf{Flow scheduling:} Given that packets reach destinations directly, flow scheduling dominates in determining per-flow performance. We consider that the packets of all active flows at the end-host are scheduled in a round-robin manner and all transmissions are paused during the synchronized reconfiguration events, similar to prior work~\cite{10.1145/3651890.3672273}. Scheduling algorithms such as shortest remaining processing time (SRPT) based on remaining flow size could potentially improve flow completion times. We leave the design of tailored scheduling algorithms for future work.

\subsection{Throughput Guarantees of Vermilion}
\label{sec:properties}

\name offers attractive theoretical guarantees on throughput and consequently completion time for a given traffic matrix. We discuss the key properties of \name in the following.

\begin{restatable}[Throughput lower bound]{theorem}{vermilionThroughputTheorem}\label{th:vermilion-throughput}
\name achieves a throughput of at least $\frac{k-1}{k}\cdot(1-\Delta_r)$ using single-hop routing for any given traffic matrix within the hose model, where $\Delta_r$ is the fraction of time spent in reconfiguration and $k$ is a parameter to \name.
\end{restatable}

\begin{proof}
We assume, without loss of generality, that the capacity of each physical link is $1$ and the degree of the physical topology is $\hat{d}$.
A traffic matrix $\mathcal{M}$ in the hose model has the property that the sum of every row and column is at most $\hat{d}$.
Let the period (number of timeslots) of the periodic schedule be $k \cdot \frac{n}{\hat{d}}$, i.e., $k$ times the period of an oblivious periodic schedule.
The emulated multigraph $G$ has a degree of $k \cdot n$, and each link has capacity $\frac{1}{k \cdot \frac{n}{\hat{d}}} = \frac{\hat{d}}{k \cdot n}$.
We now upscale both the traffic matrix $\mathcal{M}$ and each link capacity of the emulated multigraph by $\frac{k \cdot n}{\hat{d}}$. Thus, it is equivalent to find the throughput of the emulated multigraph $G$ with degree $k \cdot n$ and each link having capacity $1$, under the scaled traffic matrix $\mathcal{M}^{\prime} = \frac{k \cdot n}{\hat{d}} \cdot \mathcal{M}$, where the sum of each row and column is at most $k \cdot n$.
We seek to find the edge multiset of the emulated multigraph $G$ that maximizes throughput. In order to show that \name achieves a throughput of at least $\frac{k-1}{k}$, it suffices to show that \name can satisfy $\frac{k-1}{k} \cdot \mathcal{M}^{\prime}$ within the capacity constraints.
According to \name, edges are added based on matrix rounding using $\frac{(k-1) \cdot n}{\hat{d}} \mathcal{M} = \frac{k-1}{k} \cdot \mathcal{M}^{\prime}$. These edges satisfy all the demand in the traffic matrix $\frac{k-1}{k} \cdot \mathcal{M}^{\prime}$ except for entries that were rounded down during the rounding process. This is because, the sum of every row and column in the traffic matrix (and the rounded matrix) is at most $(k-1) \cdot n$, entries are either rounder up or down, and links are added only to non-zero entries.
The rounding process utilizes at most $(k-1) \cdot n$ incoming and outgoing links from each node. We are left with at least $n$ incoming and outgoing links for each node. The residual demand is then served by adding additional edges between each node pair.
The residual demand between any node pair is strictly less than $1$ (due to rounding), and a single additional link between the pair can fully satisfy the residual demand. Finally, each link loses $(1-\Delta_r)$ fraction of capacity due to reconfigurations and hence the overall throughput is $\frac{k-1}{k}\cdot (1-\Delta_r)$.
\end{proof}

Theorem~\ref{th:vermilion-throughput} suggests that a throughput of $\frac{k-1}{k}$ is achievable for any given traffic matrix, with $k$ acting as a control parameter for throughput. Essentially, $k$ represents the factor by which the schedule of \name is elongated compared to an oblivious schedule that provides periodic any-to-any connectivity. For instance, with $k=3$, \name guarantees a throughput of $\frac{2}{3}$, and this can be further increased by increasing $k$, resulting in higher throughput. However, this comes at the cost of longer schedules and increased delay. As a result, the throughput guarantee can only be achieved over extended periods of time, making higher $k$ values potentially unsuitable for workloads with stringent latency requirements. In our evaluations, we use $k=3$ by default, as it strikes a good balance between throughput and schedule length.

\subsection{Practicality of Vermilion}
\label{sec:practicality}

We discuss the practicality of \name in the context of modern datacenter infrastructure and optical circuit-switching technologies. A detailed discussion appears in Appendix~\ref{sec:faq}.

\medskip
\noindent
\textbf{Scalability:} 
Periodic circuit-switched networks in general exhibit excellent scalability properties~\cite{10.1145/3651890.3672248,10.1145/3098822.3098838,10.1145/3387514.3406221,Mellette:24}. The required switch size is arguably the most important scalability concern in these networks. A simple leaf-spine topology may quickly become limited in terms of the required switch size. \name can scale to large topologies with thousands of nodes interconnected by circuit switches arranged in a non-oversubscribed $k$-ary fattree topology~\cite{10.1145/1402958.1402967}. Fattree allows any permutation to be executed. For example, even a greedy algorithm can find edge-disjoint paths corresponding to a required matching. These edge-disjoint paths then reveal the required circuits at each switch in the network, consequently, the sequence of matchings for each switch can be obtained. Further, \name is much more scalable than oblivious periodic networks in terms of protocol stack as it only requires single-hop routing and does not rely on complex in-network congestion control algorithms.

\medskip
\noindent
\textbf{Complexity:}
Deriving a schedule based on \name is solvable in polynomial time. Specifically, matrix rounding is polynomial time solvable~\cite{bacharach1966matrix}, and all other transformations have a complexity of $O(n^2)$, which is inherent to any traffic-aware approach due to the need to traverse the traffic matrix at least once. We present the absolute times required to compute \name's schedule in Appendix~\ref{sec:faq}. In contrast, approaches that rely on Birkhoff decomposition not only produce schedules with variable reconfiguration durations but also face the challenge that finding a schedule of minimum length using Birkhoff decomposition is known to be NP-hard~\cite{minimumBirkhoff}. We further discuss the challenges of deriving fixed-duration schedules using Birkhoff decomposition and time quantization in Appendix~\ref{sec:faq}.

\medskip
\noindent
\textbf{Updating the schedule:}
\name being a traffic-aware design, it requires that the switches can be updated with a new periodic schedule when the communication patterns change. Efficiently updating the circuit-switching schedule is an active area of research, particularly for fast-reconfigurable periodic circuit switches operating on microsecond ($\mu$s) or nanosecond ($ns$) timescales~\cite{10.1145/3651890.3672273,7769186}. We present a detailed discussion on the various choices for switching fabrics in Appendix~\ref{sec:faq}. Commercially available optical switches that can reconfigure at millisecond scale already allow for updating the switch with arbitrary matchings via control plane~\cite{polatis}.

\medskip
\noindent
\textbf{Traffic matrix estimation:}
Modern datacenters are capable of accurately estimating the traffic matrix at scale~\cite{10.1145/3544216.3544265}. Further, more recent distributed training workloads in GPU clusters have a predictable traffic matrix that is also periodic in nature~\cite{285119,10.1145/3663408.3663409,10664412}. In Appendix~\ref{sec:faq}, we describe our approach for estimating the traffic matrix in a fully distributed manner, leveraging the oblivious phase of \name's schedule to perform an allGather operation to estimate the traffic matrix.

Overall, we believe \name does not fundamentally require novel hardware components, and is well within the practical capabilities of existing technologies.

\begin{figure*}
\centering
\begin{subfigure}{1\linewidth}
\centering
\includegraphics[width=0.7\linewidth]{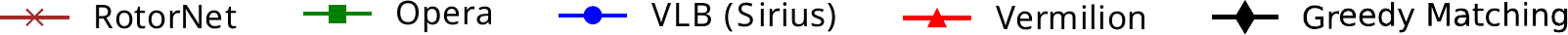}
\end{subfigure}
\begin{subfigure}{0.19\linewidth}
\centering
\includegraphics[width=1\linewidth]{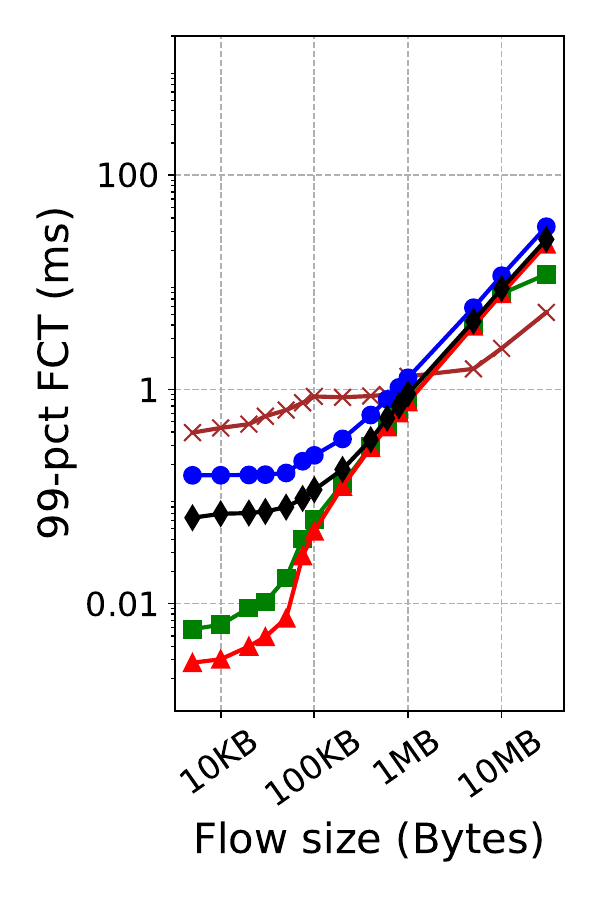}
\caption{Load=$5\%$}
\label{fig:fct-5}
\end{subfigure}\hfill
\begin{subfigure}{0.19\linewidth}
\centering
\includegraphics[width=1\linewidth]{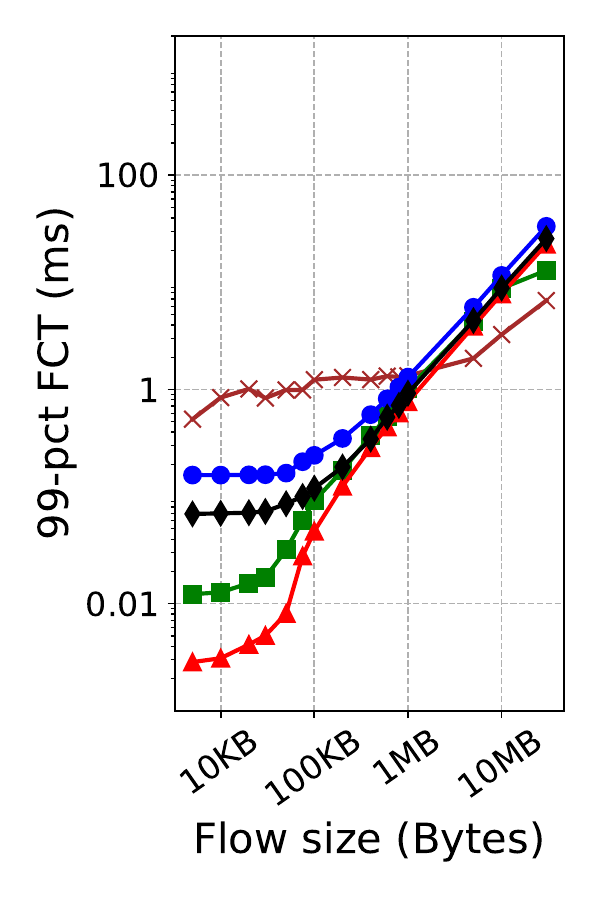}
\caption{Load=$10\%$}
\label{fig:fct-10}
\end{subfigure}\hfill
\begin{subfigure}{0.19\linewidth}
\centering
\includegraphics[width=1\linewidth]{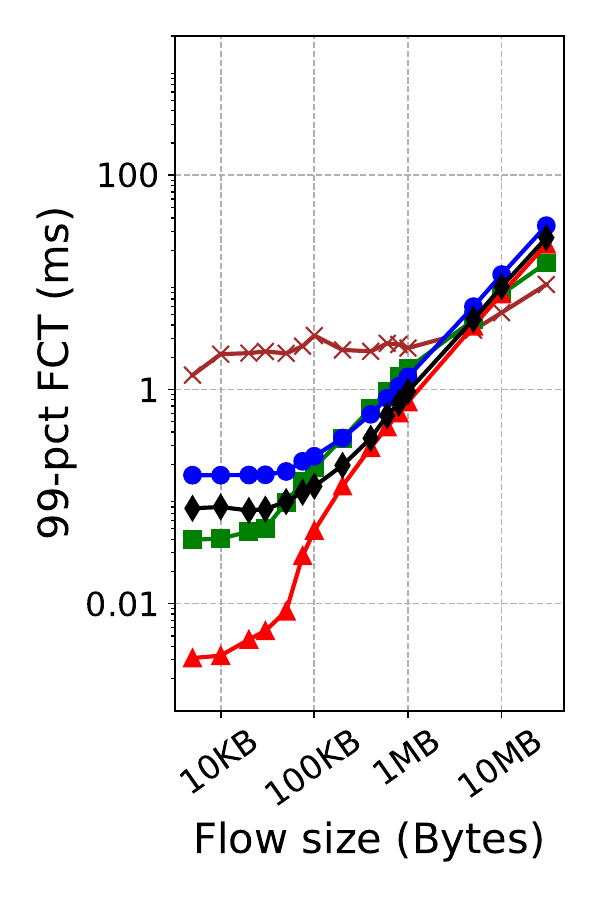}
\caption{Load=$20\%$}
\label{fig:fct-20}
\end{subfigure}\hfill
\begin{subfigure}{0.19\linewidth}
\centering
\includegraphics[width=1\linewidth]{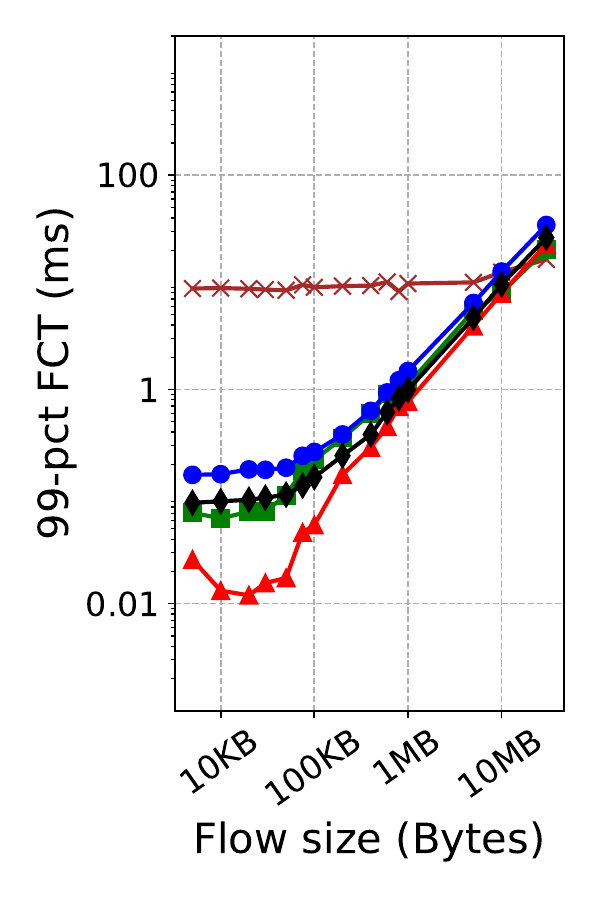}
\caption{Load=$40\%$}
\label{fig:fct-40}
\end{subfigure}\hfill
\begin{subfigure}{0.19\linewidth}
\centering
\includegraphics[width=1\linewidth]{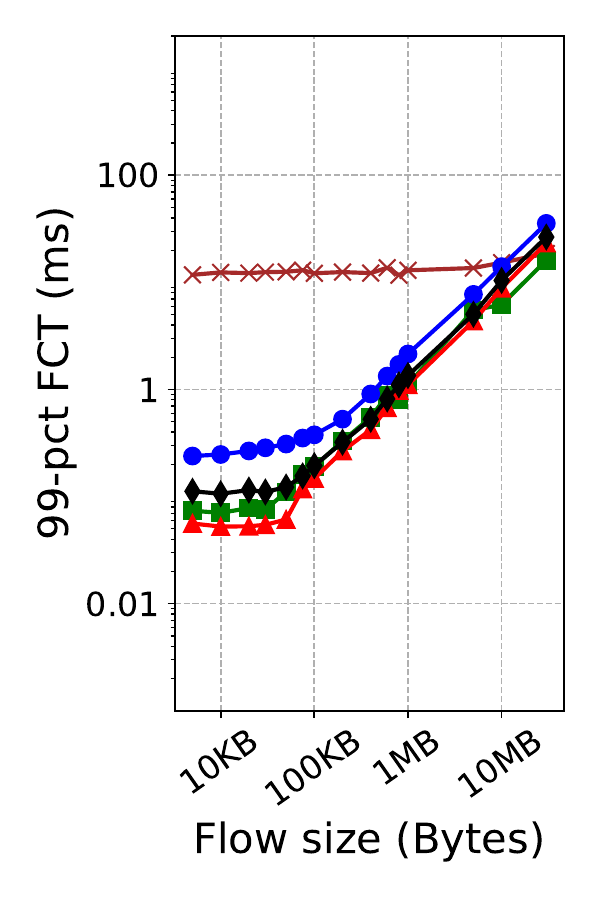}
\caption{Load=$60\%$}
\label{fig:fct-60}
\end{subfigure}\hfill
\caption{Flow completion times for the websearch workload. \name significantly improves the $99$-percentile FCTs compared to existing designs, both for short flows and long flows.}
\label{fig:fct}
\end{figure*}

\begin{figure}[t]
\centering
\begin{subfigure}{0.49\linewidth}
\includegraphics[width=1\linewidth]{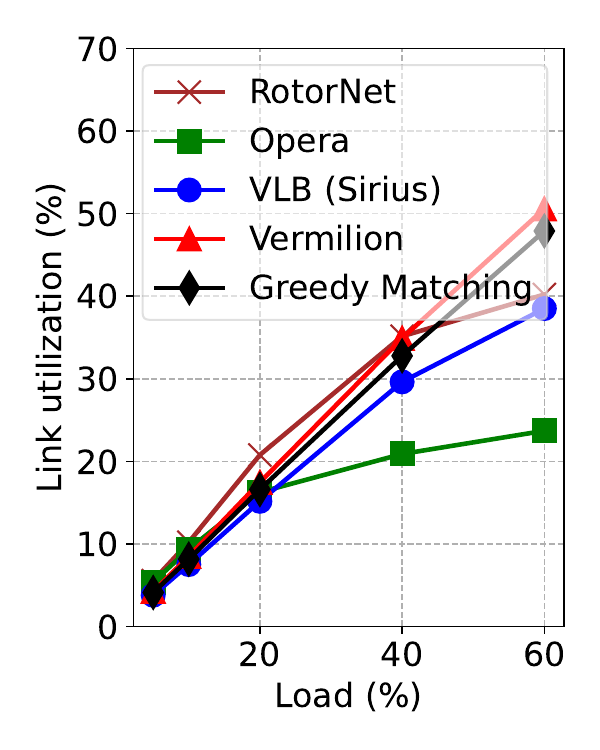}
\caption{Link utilization}
\label{fig:throughput-websearch}
\end{subfigure}
\begin{subfigure}{0.49\linewidth}
\includegraphics[width=1\linewidth]{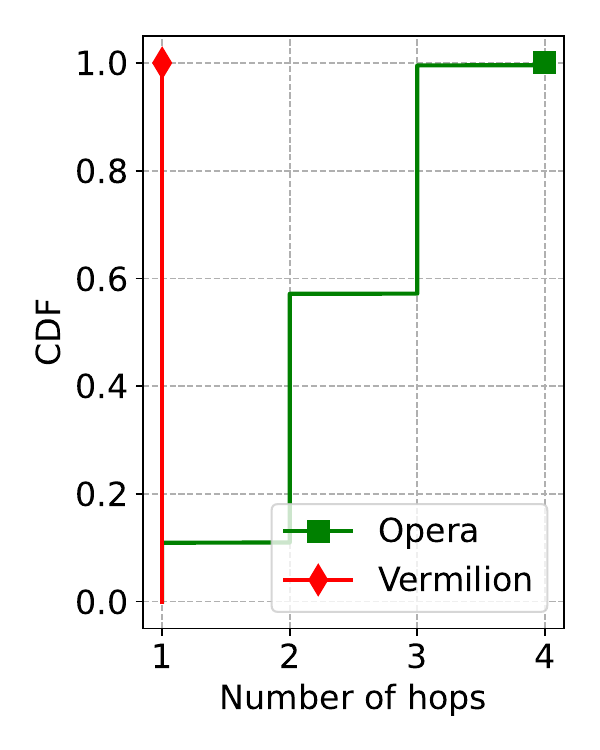}
\caption{Path lengths}
\label{fig:paths}
\end{subfigure}
\caption{Opera uses $k$-shortest paths for short flows, leading to longer path lengths and loss in link utilization, whereas \name uses exclusively direct paths  and achieves higher link utilization compared to other designs.}
\label{fig:th}
\end{figure}

\section{Evaluation}
\label{sec:evaluation}
We evaluate the performance of \name in terms of flow completion times and link utilization using real-world datacenter workloads (\S\ref{sec:htsim}), and in terms of throughput with commonly observed demand patterns in datacenters (\S\ref{sec:gurobi}). We compare \name with existing periodic network designs, namely, RotorNet~\cite{Mellette:24}, Opera~\cite{opera}, Sirius~\cite{10.1145/3387514.3406221} and a traffic-aware greedy matching baseline.

\subsection{Completion Times \& Link Utilization}
\label{sec:htsim}
Our evaluation in this section is based on packet-level simulations using htsim~\cite{opera-sim}. 

\medskip
\noindent
\textbf{Topology:}
We consider a datacenter consisting of $512$ servers arranged across $64$ top-of-rack (ToR) switches. These ToR switches are interconnected by a layer of $8$ optical circuit switches, with all link capacities set to $100$Gbps. For all systems compared, including \name, the circuit switches have a reconfiguration delay of $0.5\mu s$, which is the best-case switching time for the latest version of RotorNet~\cite{10.1145/3651890.3672273}.

\medskip
\noindent
\textbf{Comparisons \& Configurations:} We set $k=3$ for \name, and compare it with RotorNet~\cite{10.1145/3098822.3098838}, Opera~\cite{opera} and Sirius~\cite{10.1145/3387514.3406221}, representing traffic-oblivious approaches. We also compare \name with a traffic-aware approach that adapts its switching using maximum weight matching based on the underlying traffic matrix, similar to Negotiator~\cite{10.1145/3651890.3672222}. We refer to this baseline as Greedy Matching. For all systems, we set the slot time to $9\times$ the reconfiguration delay. Opera, on the other hand, internally determines its slot time based on the propagation delay~\cite{opera}. We use the recommended configurations of these systems as provided in their respective papers. Specifically, RotorNet uses RotorLB load-balancing algorithm for managing congestion between the ToRs; Opera uses $k$-shortest paths for short flows within the same timeslot to reduce FCTs; Sirius uses valiant load balancing (VLB)~\cite{valiant1982scheme,10.1145/3387514.3406221}; and \name uses single-hop routing without any further congestion control mechanisms. All systems use NDP~\cite{10.1145/3098822.3098825} as the transport protocol. In the case of \name, we turn off all actions of NDP and set a constant congestion window size.

\begin{figure*}
\centering
\begin{subfigure}{1\linewidth}
\centering
\includegraphics[width=0.8\linewidth]{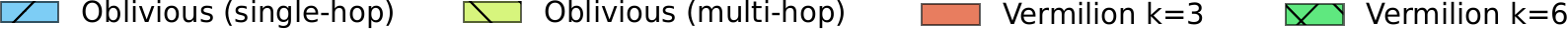}
\end{subfigure}
\begin{subfigure}{1\linewidth}
\centering
\includegraphics[width=0.9\linewidth]{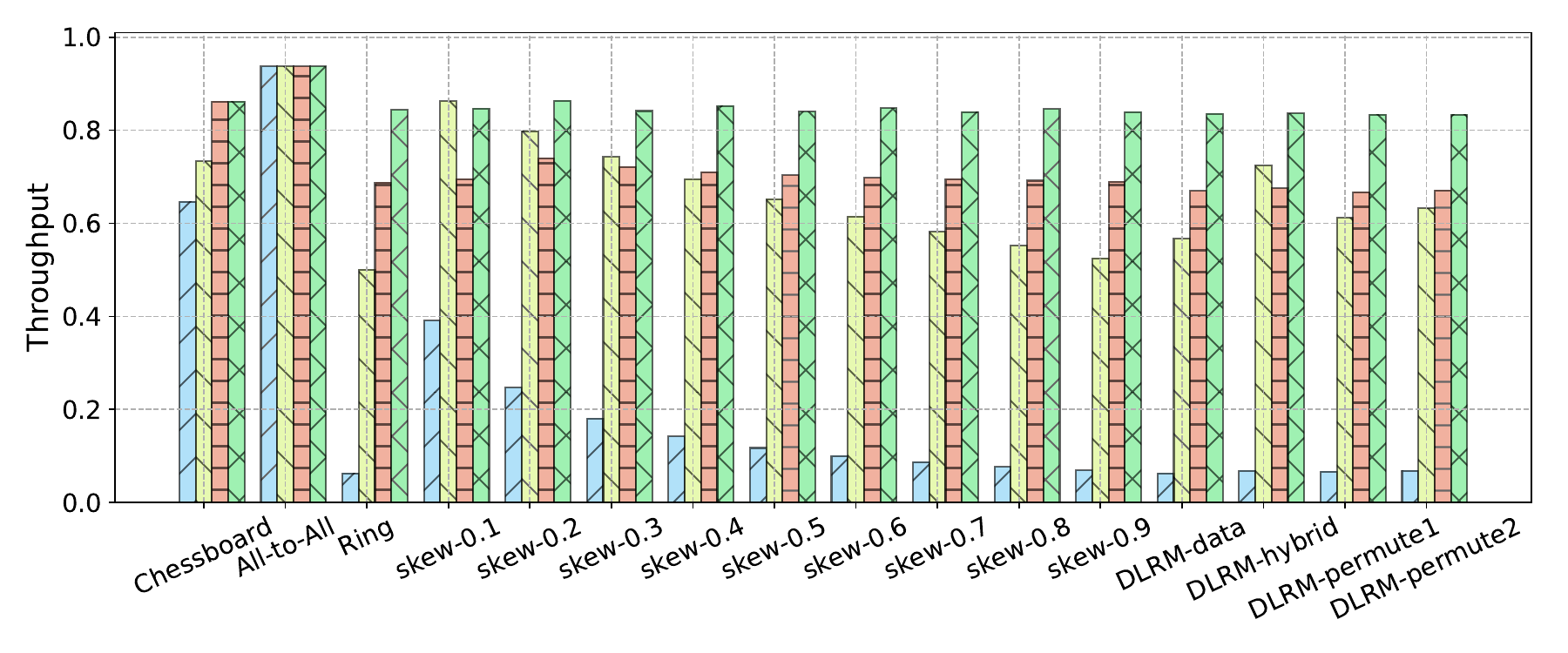}
\end{subfigure}
\caption{\name achieves higher throughput even with single-hop direct communication compared to oblivious networks with multi-hop routing, across real-world demand matrices as well as synthetic demands. The throughput of oblivious periodic networks is severely low when restricted to single-hop routing. 
}
\label{fig:throughput-16}
\end{figure*}

\begin{figure}[t]
\centering
\begin{subfigure}{0.49\linewidth}
\centering
\includegraphics[width=1\linewidth]{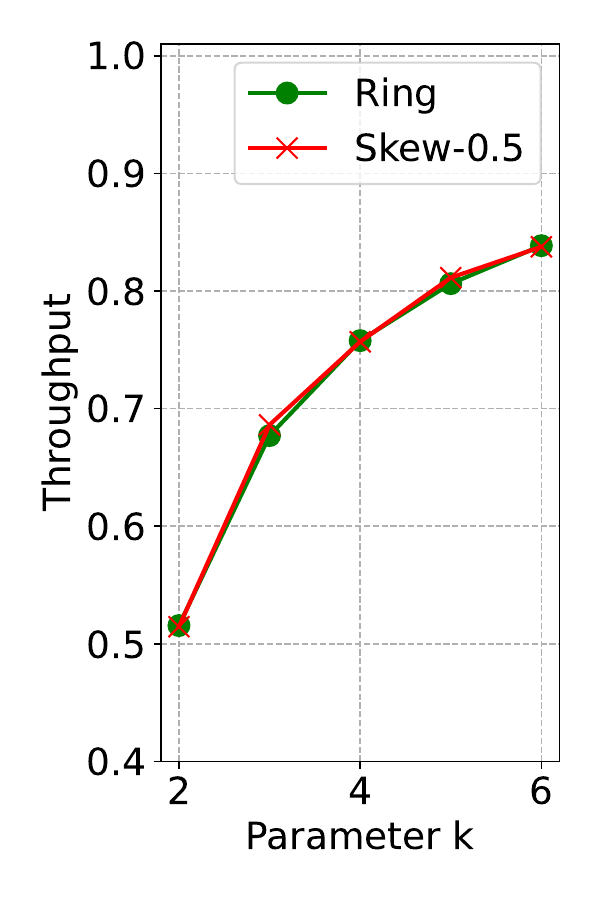}
\caption{Impact of parameter $k$}
\label{fig:throughput-k}
\end{subfigure}
\begin{subfigure}{0.495\linewidth}
\centering
\includegraphics[width=1\linewidth]{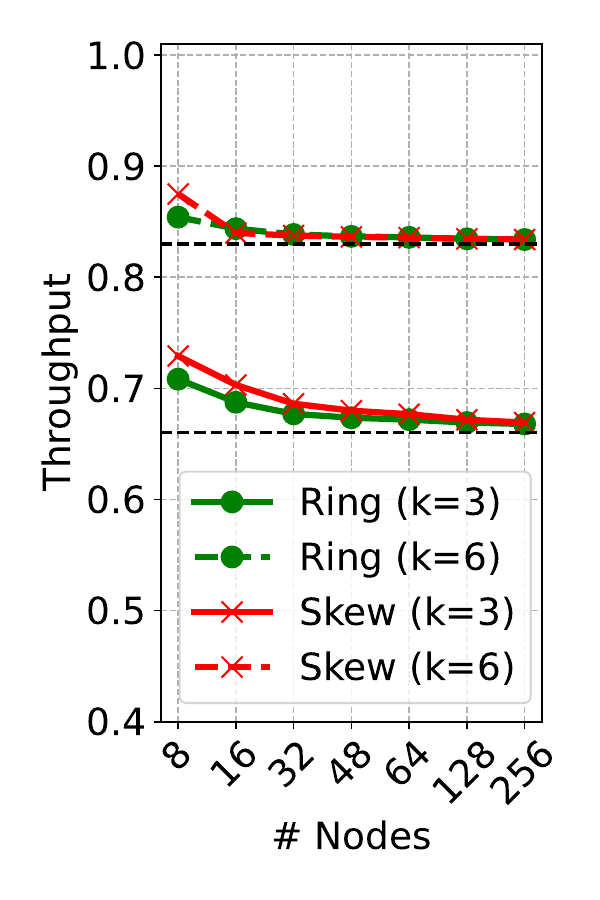}
\caption{Variation with network size}
\label{fig:throughput-nodes}
\end{subfigure}
\caption{\name's throughput follows its theoretical lower bound with increasing $k$ and converges to the bound as the network size increases.}
\label{fig:throughput-variation}
\end{figure}

\medskip
\noindent
\textbf{Workload:}
We launch the websearch~\cite{10.1145/1851182.1851192} workload, a widely-used datacenter benchmark from prior work. We simulate a pair-wise communication pattern between racks and vary the load between $5\%$ and $60\%$ of the server link capacity. Flows arrive according to a poisson process such that an average load is achieved on each server's outgoing link. 
We construct a periodic schedule for \name based on the average load\footnote{The instantaneous traffic matrix is in fact variable but we optimize based on the average load.}, while the other systems use an oblivious schedule (emulating an all-to-all mesh topology)~\cite{10.1145/3098822.3098838,opera,10.1145/3387514.3406221} for all loads. Greedy Matching baseline determines its topology based on the maximum weight matching of the underlying traffic matrix.
We report the $99$-percentile flow completion times (FCTs) and the average link utilization across server downlinks.

\medskip
\noindent
\textbf{\protect{\large{\name}} significantly improves short flow FCTs:}
Figure~\ref{fig:fct} shows the flow completion times for the websearch workload. \name significantly improves the $99$-percentile FCTs compared to existing designs, even at low loads. Figure~\ref{fig:fct-5}, at $5\%$ load, shows that \name improves the $99$-percentile FCTs by $99.28\%$ compared to RotorNet, by $51.38\%$ compared to Opera, by $98.2\%$ compared to VLB, and by $95.5\%$ compared to Greedy Matching. This is because \name provides direct communication links between communicating pairs, reducing the FCTs for short flows. As the load increases, \name improves the FCTs for short flows on average by $99.54\%$ compared to RotorNet, by $61.5\%$ compared to Opera, by $90.96\%$ compared to VLB, and by $81.57\%$ compared to Greedy Matching.

\medskip
\noindent
\textbf{\protect{\large{\name}} does not compromise long flow FCTs:}
While \name significantly reduces short flow FCTs, it also improves the FCTs for long flows. Figure~\ref{fig:fct} shows that \name outperforms alternative approaches in terms of long flow FCTs. This is primarily due to \name's traffic-aware schedule, which provides direct links between communicating pairs.
Across various loads (Figure~\ref{fig:fct}), \name achieves comparable FCTs to other systems. \name's high-throughput capability is especially beneficial for long flows, as they are bandwidth-intensive and require a robust interconnect to maintain low FCTs, while not penalizing short flows. In contrast, RotorNet achieves better FCTs for long flows, but at the cost of significantly higher FCTs for short flows.

\medskip
\noindent
\textbf{\protect{\large{\name}} improves average link utilization:}
We sample the link utilization of each server at $10\mu s$ intervals and report the average link utilization across all downlinks in the network. Figure~\ref{fig:th} highlights significant improvements in link utilization with \name. Up to $10\%$ load, \name achieves similar link utilization to RotorNet and Opera. However, as the load increases, Opera's utilization saturates at $\approx 23\%$. At $60\%$ load, \name improves average link utilization by $1.25\times$ compared to RotorNet, $2.13\times$ compared to Opera, $1.31\times$ compared to VLB, and by $1.05\times$ compared to Greedy Matching. Figure~\ref{fig:paths} presents the CDF of the number of hops taken by Opera in the ToR-to-ToR optical interconnect. Opera employs $k$-shortest paths to opportunistically reduce FCTs for short flows, but this results in longer paths, lower throughput, and consequently, reduced link utilization. In contrast, \name provides direct single-hop communication between ToR pairs, significantly enhancing link utilization.

\subsection{Throughput}
\label{sec:gurobi} 
We now evaluate the throughput capabilities of \name in comparison to existing approaches by directly analyzing throughput using a linear programming approach, eliminating protocol-level interference. We use Gurobi~\cite{gurobi} to solve the linear program for throughput maximization.

\medskip
\noindent
\textbf{Topology:} We consider a $16$ node topology with degree $4$ (incoming and outgoing links), interconnected by a layer of optical circuit switches. We set the link capacity to $25$Gbps and the reconfiguration delay to $0.5\mu s$ as before. We present our results for a larger network with $48$ nodes in Appendix~\ref{app:additional-results}.

\myitem{Demand matrices:} We evaluate across a variety of demand matrices gathered from a $16$ node GPU cluster running distributed training workload of a deep learning recommendation model, under data-parallelism, hybrid parallelism and permutations of the data-parallel workload.
We further consider synthetic demand matrices to stress the throughput capabilities of each system; parametrized by a skew parameter that combines a permutation matrix with an all-to-all uniform matrix. For instance, skew-$0.1$ indicates a $10\%$ skew towards a permutation matrix.

\medskip
\noindent
\textbf{Comparisons:} We compare \name with an ideal oblivious periodic network that emulates an all-to-all mesh topology, using an ideal routing algorithm that maximizes throughput. We call this system \textit{Oblivious (multi-hop)}. We also compare \name with oblivious systems restricted to single-hop. \name explicitly uses single-hop routing and we compare $k=3$ and $k=6$.

\medskip
\noindent
\textbf{\protect{\large{\name}} consistently achieves high throughput:}
From Figure~\ref{fig:throughput-16}, we see that \name achieves high throughput across a wide range of communication patterns. Specifically for distributed training workloads, we see that \name achieves $6.64\%$ better throughput compared to oblivious periodic networks using an ideal multi-hop routing. Figure~\ref{fig:throughput-16} shows the clear advantage (and the need) for multi-hop routing in the case of oblivious networks, with significantly lower throughput under single-hop routing. \name, on the other hand, achieves high throughput with single-hop routing. Oblivious networks with an ideal routing scheme, however, outperform \name when the traffic matrix is close to uniform. This is expected since with $k=3$, \name's lower bound is $\frac{2}{3}$. As the skew increase, oblivious network design drops to a throughput of $\frac{1}{2}$ as discussed in \S\ref{sec:drawbacks}. In contrast, \name maintains a throughput greater than $\frac{2}{3}$ even with skewed demand matrices.

\medskip
\noindent
\textbf{\protect{\large{\name}}'s throughput converges to the lower bound:}
Figure~\ref{fig:throughput-variation} confirms our theoretical bounds established in \S\ref{sec:properties}. With increasing $k$, Figure~\ref{fig:throughput-k} shows that \name's throughput closely tracks its lower bound of $\frac{k-1}{k}$. Further, even with increasing size of the network, Figure~\ref{fig:throughput-nodes} shows that \name's throughput gradually converges to the theoretical lower bound of $\frac{2}{3}$ for $k=3$ and $\frac{5}{6}$ for $k=6$, respectively. This demonstrates the robustness of \name's throughput guarantees across different network sizes and demand matrices.

\section{Limitations and Future Work}
\label{sec:future}

\name represents an initial step toward traffic-aware periodic networks capable of achieving high throughput for any traffic matrix using only single-hop routing. However, several challenges remain and open avenues for future research.

\medskip
\noindent
\textbf{Temporal dependencies in communication patterns:}
\name assumes that the traffic matrix, whether defined by rate (bits per second) or volume (bits), is available and that the demands between source-destination pairs are independent. However, certain workloads, such as distributed training, exhibit temporal dependencies in their communication patterns. For example, while the traffic matrix may accurately represent traffic, specific portions of the demand (e.g., from the backward pass) may only become available after the completion of other parts (e.g., from the forward pass)~\cite{10.1145/3663408.3663409,10664412,10.1145/3651890.3672249}. Addressing these temporal dependencies presents a significant research challenge and offers opportunities to further optimize topologies like \name for such workloads.

\medskip
\noindent
\textbf{Structured communication patterns:}
\name is designed to achieve high throughput for any arbitrary traffic matrix. However, some communication patterns have inherent structure that could be exploited for further optimization. For example, the ring-allReduce collective communication, commonly used in distributed training, can be efficiently supported by a simple ring-emulated topology with periodic schedules. Optimizing for specific communication patterns is complementary to our approach. In principle, \name could be extended to recognize and leverage these structured communication patterns using existing solutions~\cite{efficientdirectnsdi2025}, potentially achieving even higher throughput and faster completion times. Exploring these optimizations is an avenue for future work.

\medskip
\noindent
\textbf{Fault-tolerance and resilience:}
\name does not explicitly address fault-tolerance or resilience. While the periodic nature of the network may offer some inherent resilience to failures, designing fault-tolerant periodic schedules remains an open challenge. Failures in optical networks can be particularly difficult to detect, as they often manifest as packet corruption due to optical collisions. Recent work discusses techniques to mitigate link-layer and physical-layer errors~\cite{10.1145/3651890.3672273}. Future work could explore the design of fault-tolerant schedules that can quickly adapt to failures while still maintaining high throughput.

\medskip
\noindent
\textbf{Heterogeneous link capacities:}
In this paper, we assume that all physical links in the topology have uniform capacity. However, datacenter topologies often include links with heterogeneous capacities. For example, faulty auto-negotiation between two NICs can result in a link operating at a lower capacity than expected. In practice, these capacities are often multiples of a base rate. We believe \name can be generalized to accommodate heterogeneous link capacities by selecting an appropriate base capacity and adjusting the topology to handle varying multiples of that capacity. We leave the generalization of \name for heterogeneous link capacities to future work.

\section{Related Work}
\label{sec:related}

Datacenter topologies have been widely studied in the literature both in the context of traditional packet-switched networks~\cite{10.1145/2999572.2999580,180604,10.1145/1402958.1402967,10.1145/1592568.1592576,227667,10.1145/2785956.2787508,7013016,f10,10.1145/1592568.1592577} and emerging reconfigurable optically circuit-switched networks~\cite{10.1145/3098822.3098838,10.1145/3387514.3406221,10.1145/2934872.2934911,10.1145/1851182.1851223,10.1145/3579449,10.1145/2377677.2377761,kandula2009flyways,opera,10.1145/2619239.2626328,201560,10.1145/2619239.2626332,6490069,10.1145/1851182.1851222,7066977,10.1145/2896377.2901479,10.1145/1868447.1868455,10.1145/3491050,278374,10.1145/3351452.3351464}. In the design of topologies, various metrics of interest have been considered. For instance, uniformly high bandwidth availability~\cite{10.1145/1402958.1402967,10.1145/1592568.1592577}, expansion~\cite{10.1145/2999572.2999580,180604}, fault-tolerance~\cite{f10}, and even the life cycle management of a datacenter~\cite{227667}. In the context of reconfigurable networks, typically, the goal has been either to minimize the reconfiguration overhead~\cite{10.1145/3098822.3098838,10.1145/3387514.3406221} or to minimize the bandwidth tax~\cite{10.1145/2934872.2934911,10.1145/1851182.1851223,10.1145/3579449,285119}.

Recent works argue for a new measure \ie ``throughput'', to understand the maximum load supported by a topology~\cite{7877143,179775,10.1145/3452296.3472913,10.1145/3491050,10.1145/3579312}. In fact, the max-flow that relates to the throughput of a topology, can be $\mathcal{O}(\log n)$ factor lower than the sparsest cut~\cite{10.1145/331524.331526,10.1109/SFCS.1988.21958,7877143}. Namyar et al.~study the throughput upper bound for static datacenter topologies and show a separation between Clos (i.e., fat-trees) and expander-based networks in terms of throughput~\cite{10.1145/3452296.3472913}. In the context of reconfigurable networks, only recently have the throughput bounds of traffic-oblivious networks been established~\cite{10.1145/3579312,10.1145/3491050,10.1145/3519935.3520020}. 

While throughput of a datacenter topology is interesting from a theory standpoint, a vast majority of the literature focuses on practically achieving the ideal throughput of a topology. For instance, congestion control~\cite{10.1145/1851182.1851192,10.1145/2785956.2787510,10.1145/3341302.3342085,278346,10.1145/3387514.3406591,276958,10.1145/3387514.3405899,10.1145/2018436.2018443,uec}, buffer management~\cite{abm,fab,trafficaware,295539,10229046,295535}, scheduling~\cite{10.1145/2486001.2486031,259355,cassini}, load-balancing~\cite{10.1145/2619239.2626316,10.1145/2890955.2890968,10.1145/3098822.3098839,bonato2025repsrecycledentropypacket,addanki2025etherealdivideconquernetwork}. In fact, the underlying protocols can turn out to be the key enablers (or limiters) of system performance in the datacenter~\cite{10.1145/3387514.3406591}. Only recently, congestion control~\cite{10.1145/3603269.3610862,10.1145/3544216.3544254,246336} and routing~\cite{10.1145/3651890.3672245} algorithms tailored for reconfigurable networks have been considered. 
Interestingly, if \name is deployed in an all-optical setting, it does not require any additional congestion control, buffer management and load-balancing mechanisms, since it relies solely on direct communication. 

\section{Conclusion}
\label{sec:conclusion}
We introduced \name, a simple traffic-aware optical interconnect that achieves high throughput using only periodic circuit-switching and direct communication. Through formal analysis, we established throughput bounds for \name, marking the first formal separation result that demonstrates traffic-aware reconfigurable networks' superiority over oblivious counterparts in terms of throughput. We believe that \name offers a practical solution for datacenter networks with predictable communication patterns. In the future, we plan to explore the temporal dependencies in communication patterns that arise in distributed training workloads and investigate how \name can be further optimized for such~scenarios.

\section*{Acknowledgments}

\noindent
This work is part of a project that has received funding from the European Research Council (ERC) under the European Union's Horizon 2020 research and innovation programme, consolidator project Self-Adjusting Networks (AdjustNet), grant agreement No. 864228, Horizon 2020, 2020-2025. Chen~Avin was additionally partially supported by the Israeli Science Foundation, grant ISF 2497/23.
\begin{figure}[!h]
    \centering
    \includegraphics[width=0.6\linewidth]{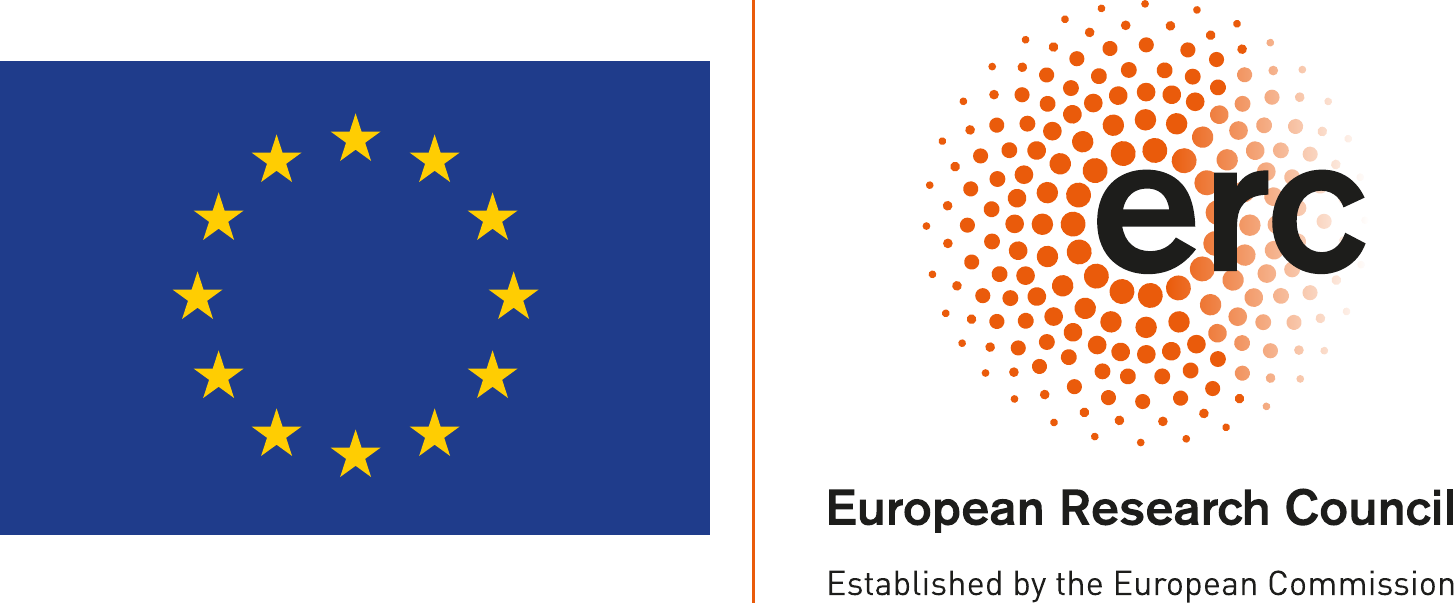}
    \label{fig:my_label}
\end{figure}

\label{bodyLastPage}
\pagebreak
\bibliographystyle{unsrt}
\bibliography{references}

\begin{thebibliography}{10}

\bibitem{10.1145/2785956.2787508}
Arjun Singh, Joon Ong, Amit Agarwal, Glen Anderson, Ashby Armistead, Roy
  Bannon, Seb Boving, Gaurav Desai, Bob Felderman, Paulie Germano, Anand
  Kanagala, Jeff Provost, Jason Simmons, Eiichi Tanda, Jim Wanderer, Urs
  H\"{o}lzle, Stephen Stuart, and Amin Vahdat.
\newblock Jupiter rising: A decade of clos topologies and centralized control
  in google's datacenter network.
\newblock In {\em Proceedings of the 2015 ACM Conference on Special Interest
  Group on Data Communication}, SIGCOMM '15, page 183–197, New York, NY, USA,
  2015. Association for Computing Machinery.

\bibitem{10.1145/3544216.3544265}
Leon Poutievski, Omid Mashayekhi, Joon Ong, Arjun Singh, Mukarram Tariq, Rui
  Wang, Jianan Zhang, Virginia Beauregard, Patrick Conner, Steve Gribble, Rishi
  Kapoor, Stephen Kratzer, Nanfang Li, Hong Liu, Karthik Nagaraj, Jason
  Ornstein, Samir Sawhney, Ryohei Urata, Lorenzo Vicisano, Kevin Yasumura,
  Shidong Zhang, Junlan Zhou, and Amin Vahdat.
\newblock Jupiter evolving: transforming google's datacenter network via
  optical circuit switches and software-defined networking.
\newblock In {\em Proceedings of the ACM SIGCOMM 2022 Conference}, SIGCOMM '22,
  page 66–85, New York, NY, USA, 2022. Association for Computing Machinery.

\bibitem{10.1145/3387514.3406221}
Hitesh Ballani, Paolo Costa, Raphael Behrendt, Daniel Cletheroe, Istvan Haller,
  Krzysztof Jozwik, Fotini Karinou, Sophie Lange, Kai Shi, Benn Thomsen, and
  Hugh Williams.
\newblock Sirius: A flat datacenter network with nanosecond optical switching.
\newblock In {\em Proceedings of the Annual Conference of the ACM Special
  Interest Group on Data Communication on the Applications, Technologies,
  Architectures, and Protocols for Computer Communication}, SIGCOMM '20, page
  782–797, New York, NY, USA, 2020. Association for Computing Machinery.

\bibitem{10.1145/3098822.3098838}
William~M. Mellette, Rob McGuinness, Arjun Roy, Alex Forencich, George Papen,
  Alex~C. Snoeren, and George Porter.
\newblock Rotornet: A scalable, low-complexity, optical datacenter network.
\newblock In {\em Proceedings of the Conference of the ACM Special Interest
  Group on Data Communication}, SIGCOMM '17, page 267–280, New York, NY, USA,
  2017. Association for Computing Machinery.

\bibitem{10.1145/2934872.2934911}
Monia Ghobadi, Ratul Mahajan, Amar Phanishayee, Nikhil Devanur, Janardhan
  Kulkarni, Gireeja Ranade, Pierre-Alexandre Blanche, Houman Rastegarfar,
  Madeleine Glick, and Daniel Kilper.
\newblock Projector: Agile reconfigurable data center interconnect.
\newblock In {\em Proceedings of the 2016 ACM SIGCOMM Conference}, SIGCOMM '16,
  page 216–229, New York, NY, USA, 2016. Association for Computing Machinery.

\bibitem{10.1145/1851182.1851223}
Nathan Farrington, George Porter, Sivasankar Radhakrishnan, Hamid~Hajabdolali
  Bazzaz, Vikram Subramanya, Yeshaiahu Fainman, George Papen, and Amin Vahdat.
\newblock Helios: a hybrid electrical/optical switch architecture for modular
  data centers.
\newblock In {\em Proceedings of the ACM SIGCOMM 2010 Conference}, SIGCOMM '10,
  page 339–350, New York, NY, USA, 2010. Association for Computing Machinery.

\bibitem{10.1145/3579312}
Vamsi Addanki, Chen Avin, and Stefan Schmid.
\newblock Mars: Near-optimal throughput with shallow buffers in reconfigurable
  datacenter networks.
\newblock {\em Proc. ACM Meas. Anal. Comput. Syst.}, 7(1), mar 2023.

\bibitem{10.1145/3651890.3672248}
Daniel Amir, Nitika Saran, Tegan Wilson, Robert Kleinberg, Vishal Shrivastav,
  and Hakim Weatherspoon.
\newblock Shale: A practical, scalable oblivious reconfigurable network.
\newblock In {\em Proceedings of the ACM SIGCOMM 2024 Conference}, ACM SIGCOMM
  '24, page 449–464, New York, NY, USA, 2024. Association for Computing
  Machinery.

\bibitem{10.1145/3579449}
Johannes Zerwas, Csaba Gy\"{o}rgyi, Andreas Blenk, Stefan Schmid, and Chen
  Avin.
\newblock Duo: A high-throughput reconfigurable datacenter network using local
  routing and control.
\newblock {\em Proc. ACM Meas. Anal. Comput. Syst.}, 7(1), mar 2023.

\bibitem{opera}
William~M. Mellette, Rajdeep Das, Yibo Guo, Rob McGuinness, Alex~C. Snoeren,
  and George Porter.
\newblock Expanding across time to deliver bandwidth efficiency and low
  latency.
\newblock In {\em 17th USENIX Symposium on Networked Systems Design and
  Implementation (NSDI 20)}, pages 1--18, Santa Clara, CA, February 2020.
  USENIX Association.

\bibitem{10.1145/3519935.3520020}
Daniel Amir, Tegan Wilson, Vishal Shrivastav, Hakim Weatherspoon, Robert
  Kleinberg, and Rachit Agarwal.
\newblock Optimal oblivious reconfigurable networks.
\newblock In {\em Proceedings of the 54th Annual ACM SIGACT Symposium on Theory
  of Computing}, STOC 2022, page 1339–1352, New York, NY, USA, 2022.
  Association for Computing Machinery.

\bibitem{10.1145/2486001.2486007}
George Porter, Richard Strong, Nathan Farrington, Alex Forencich, Pang
  Chen-Sun, Tajana Rosing, Yeshaiahu Fainman, George Papen, and Amin Vahdat.
\newblock Integrating microsecond circuit switching into the data center.
\newblock In {\em Proceedings of the ACM SIGCOMM 2013 Conference on SIGCOMM},
  SIGCOMM '13, page 447–458, New York, NY, USA, 2013. Association for
  Computing Machinery.

\bibitem{Mellette:24}
William~M. Mellette, Ilya Agurok, Alex Forencich, Spencer Chang, George Papen,
  and Joseph~E. Ford.
\newblock A scalable, high-speed optical rotor switch.
\newblock In {\em Optical Fiber Communication Conference (OFC) 2024}, page
  Th1A.5. Optica Publishing Group, 2024.

\bibitem{10.1145/3651890.3672273}
William~M. Mellette, Alex Forencich, Rukshani Athapathu, Alex~C. Snoeren,
  George Papen, and George Porter.
\newblock Realizing rotornet: Toward practical microsecond scale optical
  networking.
\newblock In {\em Proceedings of the ACM SIGCOMM 2024 Conference}, ACM SIGCOMM
  '24, page 392–414, New York, NY, USA, 2024. Association for Computing
  Machinery.

\bibitem{10.1145/3491050}
Chen Griner, Johannes Zerwas, Andreas Blenk, Manya Ghobadi, Stefan Schmid, and
  Chen Avin.
\newblock Cerberus: The power of choices in datacenter topology design - a
  throughput perspective.
\newblock {\em Proc. ACM Meas. Anal. Comput. Syst.}, 5(3), dec 2021.

\bibitem{birkhoff1946three}
Garrett Birkhoff.
\newblock Tres observaciones sobre el algebra lineal.
\newblock {\em Univ. Nac. Tucuman, Ser. A}, 5:147--154, 1946.

\bibitem{10.1145/2716281.2836126}
He~Liu, Matthew~K. Mukerjee, Conglong Li, Nicolas Feltman, George Papen, Stefan
  Savage, Srinivasan Seshan, Geoffrey~M. Voelker, David~G. Andersen, Michael
  Kaminsky, George Porter, and Alex~C. Snoeren.
\newblock Scheduling techniques for hybrid circuit/packet networks.
\newblock In {\em Proceedings of the 11th ACM Conference on Emerging Networking
  Experiments and Technologies}, CoNEXT '15, New York, NY, USA, 2015.
  Association for Computing Machinery.

\bibitem{corundum}
Alex Forencich, Alex~C. Snoeren, George Porter, and George Papen.
\newblock Corundum: An open-source 100-gbps nic.
\newblock In {\em 2020 IEEE 28th Annual International Symposium on
  Field-Programmable Custom Computing Machines (FCCM)}, pages 38--46, 2020.

\bibitem{bacharach1966matrix}
Michael Bacharach.
\newblock Matrix rounding problems.
\newblock {\em Management Science}, 12(9):732--742, 1966.

\bibitem{10.1145/2896377.2901479}
Shaileshh Bojja~Venkatakrishnan, Mohammad Alizadeh, and Pramod Viswanath.
\newblock Costly circuits, submodular schedules and approximate
  carath\'{e}odory theorems.
\newblock In {\em Proceedings of the 2016 ACM SIGMETRICS International
  Conference on Measurement and Modeling of Computer Science}, SIGMETRICS '16,
  page 75–88, New York, NY, USA, 2016. Association for Computing Machinery.

\bibitem{10.1145/3452296.3472913}
Pooria Namyar, Sucha Supittayapornpong, Mingyang Zhang, Minlan Yu, and Ramesh
  Govindan.
\newblock A throughput-centric view of the performance of datacenter
  topologies.
\newblock In {\em Proceedings of the 2021 ACM SIGCOMM 2021 Conference}, SIGCOMM
  '21, page 349–369, New York, NY, USA, 2021. Association for Computing
  Machinery.

\bibitem{10.1145/316188.316209}
N.~G. Duffield, Pawan Goyal, Albert Greenberg, Partho Mishra, K.~K.
  Ramakrishnan, and Jacobus~E. van~der Merive.
\newblock A flexible model for resource management in virtual private networks.
\newblock In {\em Proceedings of the Conference on Applications, Technologies,
  Architectures, and Protocols for Computer Communication}, SIGCOMM '99, page
  95–108, New York, NY, USA, 1999. Association for Computing Machinery.

\bibitem{7877143}
Sangeetha~Abdu Jyothi, Ankit Singla, P.~Brighten Godfrey, and Alexandra Kolla.
\newblock Measuring and understanding throughput of network topologies.
\newblock In {\em SC '16: Proceedings of the International Conference for High
  Performance Computing, Networking, Storage and Analysis}, pages 761--772,
  2016.

\bibitem{10.1145/77600.77620}
Farhad Shahrokhi and D.~W. Matula.
\newblock The maximum concurrent flow problem.
\newblock {\em J. ACM}, 37(2):318–334, apr 1990.

\bibitem{10.1145/3409964.3461786}
Janardhan Kulkarni, Stefan Schmid, and Pawe\l{} Schmidt.
\newblock Scheduling opportunistic links in two-tiered reconfigurable
  datacenters.
\newblock In {\em Proceedings of the 33rd ACM Symposium on Parallelism in
  Algorithms and Architectures}, SPAA '21, page 318–327, New York, NY, USA,
  2021. Association for Computing Machinery.

\bibitem{10.1145/1402958.1402967}
Mohammad Al-Fares, Alexander Loukissas, and Amin Vahdat.
\newblock A scalable, commodity data center network architecture.
\newblock In {\em Proceedings of the ACM SIGCOMM 2008 Conference on Data
  Communication}, SIGCOMM '08, page 63–74, New York, NY, USA, 2008.
  Association for Computing Machinery.

\bibitem{minimumBirkhoff}
Janardhan Kulkarni, Euiwoong Lee, and Mohit Singh.
\newblock Minimum birkhoff-von neumann decomposition.
\newblock In Friedrich Eisenbrand and Jochen Koenemann, editors, {\em Integer
  Programming and Combinatorial Optimization}, pages 343--354, Cham, 2017.
  Springer International Publishing.

\bibitem{7769186}
William~Maxwell Mellette, Glenn~M. Schuster, George Porter, George Papen, and
  Joseph~E. Ford.
\newblock A scalable, partially configurable optical switch for data center
  networks.
\newblock {\em Journal of Lightwave Technology}, 35(2):136--144, 2017.

\bibitem{polatis}
{Polatis}.
\newblock {POLATIS® 7000 Optical Circuit Switch}.
\newblock
  \url{https://www.hubersuhner.com/en/shop/product/other-systems/optical-switches/rack-mount-circuit-switches/85223159/polatis-7000-optical-circuit-switch}.

\bibitem{285119}
Weiyang Wang, Moein Khazraee, Zhizhen Zhong, Manya Ghobadi, Zhihao Jia,
  Dheevatsa Mudigere, Ying Zhang, and Anthony Kewitsch.
\newblock {TopoOpt}: Co-optimizing network topology and parallelization
  strategy for distributed training jobs.
\newblock In {\em 20th USENIX Symposium on Networked Systems Design and
  Implementation (NSDI 23)}, pages 739--767, Boston, MA, April 2023. USENIX
  Association.

\bibitem{10.1145/3663408.3663409}
Wenxue Li, Xiangzhou Liu, Yuxuan Li, Yilun Jin, Han Tian, Zhizhen Zhong, Guyue
  Liu, Ying Zhang, and Kai Chen.
\newblock Understanding communication characteristics of distributed training.
\newblock In {\em Proceedings of the 8th Asia-Pacific Workshop on Networking},
  APNet '24, page 1–8, New York, NY, USA, 2024. Association for Computing
  Machinery.

\bibitem{10664412}
Weiyang Wang, Manya Ghobadi, Kayvon Shakeri, Ying Zhang, and Naader Hasani.
\newblock Rail-only: A low-cost high-performance network for training llms with
  trillion parameters.
\newblock In {\em 2024 IEEE Symposium on High-Performance Interconnects
  (HOTI)}, pages 1--10, Los Alamitos, CA, USA, aug 2024. IEEE Computer Society.

\bibitem{opera-sim}
opera sim.
\newblock \url{https://github.com/TritonNetworking/opera-sim}.

\bibitem{10.1145/3651890.3672222}
Cong Liang, Xiangli Song, Jing Cheng, Mowei Wang, Yashe Liu, Zhenhua Liu,
  Shizhen Zhao, and Yong Cui.
\newblock Negotiator: Towards a simple yet effective on-demand reconfigurable
  datacenter network.
\newblock In {\em Proceedings of the ACM SIGCOMM 2024 Conference}, ACM SIGCOMM
  '24, page 415–432, New York, NY, USA, 2024. Association for Computing
  Machinery.

\bibitem{valiant1982scheme}
Leslie~G. Valiant.
\newblock A scheme for fast parallel communication.
\newblock {\em SIAM journal on computing}, 11(2):350--361, 1982.

\bibitem{10.1145/3098822.3098825}
Mark Handley, Costin Raiciu, Alexandru Agache, Andrei Voinescu, Andrew~W.
  Moore, Gianni Antichi, and Marcin W\'{o}jcik.
\newblock Re-architecting datacenter networks and stacks for low latency and
  high performance.
\newblock In {\em Proceedings of the Conference of the ACM Special Interest
  Group on Data Communication}, SIGCOMM '17, page 29–42, New York, NY, USA,
  2017. Association for Computing Machinery.

\bibitem{10.1145/1851182.1851192}
Mohammad Alizadeh, Albert Greenberg, David~A. Maltz, Jitendra Padhye, Parveen
  Patel, Balaji Prabhakar, Sudipta Sengupta, and Murari Sridharan.
\newblock Data center tcp (dctcp).
\newblock In {\em Proceedings of the ACM SIGCOMM 2010 Conference}, SIGCOMM '10,
  page 63–74, New York, NY, USA, 2010. Association for Computing Machinery.

\bibitem{gurobi}
{Gurobi Optimization, LLC}.
\newblock {Gurobi Optimizer Reference Manual}, 2023.

\bibitem{10.1145/3651890.3672249}
Xuting Liu, Behnaz Arzani, Siva Kesava~Reddy Kakarla, Liangyu Zhao, Vincent
  Liu, Miguel Castro, Srikanth Kandula, and Luke Marshall.
\newblock Rethinking machine learning collective communication as a
  multi-commodity flow problem.
\newblock In {\em Proceedings of the ACM SIGCOMM 2024 Conference}, ACM SIGCOMM
  '24, page 16–37, New York, NY, USA, 2024. Association for Computing
  Machinery.

\bibitem{efficientdirectnsdi2025}
Liangyu Zhao, Siddharth Pal, Tapan Chugh, Weiyang Wang, Jason Fantl, Prithwish
  Basu, Joud Khoury, and Arvind Krishnamurthy.
\newblock Efficient direct-connect topologies for collective communications.
\newblock {\em CoRR}, abs/2202.03356, 2024.

\bibitem{10.1145/2999572.2999580}
Asaf Valadarsky, Gal Shahaf, Michael Dinitz, and Michael Schapira.
\newblock Xpander: Towards optimal-performance datacenters.
\newblock In {\em Proceedings of the 12th International on Conference on
  Emerging Networking EXperiments and Technologies}, CoNEXT '16, page
  205–219, New York, NY, USA, 2016. Association for Computing Machinery.

\bibitem{180604}
Ankit Singla, Chi-Yao Hong, Lucian Popa, and P.~Brighten Godfrey.
\newblock Jellyfish: Networking data centers randomly.
\newblock In {\em 9th USENIX Symposium on Networked Systems Design and
  Implementation (NSDI 12)}, pages 225--238, San Jose, CA, April 2012. USENIX
  Association.

\bibitem{10.1145/1592568.1592576}
Albert Greenberg, James~R. Hamilton, Navendu Jain, Srikanth Kandula, Changhoon
  Kim, Parantap Lahiri, David~A. Maltz, Parveen Patel, and Sudipta Sengupta.
\newblock Vl2: a scalable and flexible data center network.
\newblock In {\em Proceedings of the ACM SIGCOMM 2009 Conference on Data
  Communication}, SIGCOMM '09, page 51–62, New York, NY, USA, 2009.
  Association for Computing Machinery.

\bibitem{227667}
Mingyang Zhang, Radhika~Niranjan Mysore, Sucha Supittayapornpong, and Ramesh
  Govindan.
\newblock Understanding lifecycle management complexity of datacenter
  topologies.
\newblock In {\em 16th USENIX Symposium on Networked Systems Design and
  Implementation (NSDI 19)}, pages 235--254, Boston, MA, February 2019. USENIX
  Association.

\bibitem{7013016}
Maciej Besta and Torsten Hoefler.
\newblock Slim fly: A cost effective low-diameter network topology.
\newblock In {\em SC '14: Proceedings of the International Conference for High
  Performance Computing, Networking, Storage and Analysis}, pages 348--359,
  2014.

\bibitem{f10}
Vincent Liu, Daniel Halperin, Arvind Krishnamurthy, and Thomas Anderson.
\newblock F10: A {{Fault-Tolerant}} engineered network.
\newblock In {\em 10th USENIX Symposium on Networked Systems Design and
  Implementation (NSDI 13)}, pages 399--412, Lombard, IL, April 2013. USENIX
  Association.

\bibitem{10.1145/1592568.1592577}
Chuanxiong Guo, Guohan Lu, Dan Li, Haitao Wu, Xuan Zhang, Yunfeng Shi, Chen
  Tian, Yongguang Zhang, and Songwu Lu.
\newblock Bcube: a high performance, server-centric network architecture for
  modular data centers.
\newblock In {\em Proceedings of the ACM SIGCOMM 2009 Conference on Data
  Communication}, SIGCOMM '09, page 63–74, New York, NY, USA, 2009.
  Association for Computing Machinery.

\bibitem{10.1145/2377677.2377761}
Xia Zhou, Zengbin Zhang, Yibo Zhu, Yubo Li, Saipriya Kumar, Amin Vahdat, Ben~Y.
  Zhao, and Haitao Zheng.
\newblock Mirror mirror on the ceiling: flexible wireless links for data
  centers.
\newblock {\em SIGCOMM Comput. Commun. Rev.}, 42(4):443–454, aug 2012.

\bibitem{kandula2009flyways}
Srikanth Kandula, Jitendra Padhye, and Paramvir Bahl.
\newblock Flyways to de-congest data center networks.
\newblock In {\em HotNets}. {ACM} {SIGCOMM}, 2009.

\bibitem{10.1145/2619239.2626328}
Navid Hamedazimi, Zafar Qazi, Himanshu Gupta, Vyas Sekar, Samir~R. Das, Jon~P.
  Longtin, Himanshu Shah, and Ashish Tanwer.
\newblock Firefly: a reconfigurable wireless data center fabric using
  free-space optics.
\newblock In {\em Proceedings of the 2014 ACM Conference on SIGCOMM}, SIGCOMM
  '14, page 319–330, New York, NY, USA, 2014. Association for Computing
  Machinery.

\bibitem{201560}
Li~Chen, Kai Chen, Zhonghua Zhu, Minlan Yu, George Porter, Chunming Qiao, and
  Shan Zhong.
\newblock Enabling {Wide-Spread} communications on optical fabric with
  {MegaSwitch}.
\newblock In {\em 14th USENIX Symposium on Networked Systems Design and
  Implementation (NSDI 17)}, pages 577--593, Boston, MA, March 2017. USENIX
  Association.

\bibitem{10.1145/2619239.2626332}
Yunpeng~James Liu, Peter~Xiang Gao, Bernard Wong, and Srinivasan Keshav.
\newblock Quartz: a new design element for low-latency dcns.
\newblock In {\em Proceedings of the 2014 ACM Conference on SIGCOMM}, SIGCOMM
  '14, page 283–294, New York, NY, USA, 2014. Association for Computing
  Machinery.

\bibitem{6490069}
Kai Chen, Ankit Singla, Atul Singh, Kishore Ramachandran, Lei Xu, Yueping
  Zhang, Xitao Wen, and Yan Chen.
\newblock Osa: An optical switching architecture for data center networks with
  unprecedented flexibility.
\newblock {\em IEEE/ACM Transactions on Networking}, 22(2):498--511, 2014.

\bibitem{10.1145/1851182.1851222}
Guohui Wang, David~G. Andersen, Michael Kaminsky, Konstantina Papagiannaki,
  T.S.~Eugene Ng, Michael Kozuch, and Michael Ryan.
\newblock c-through: part-time optics in data centers.
\newblock In {\em Proceedings of the ACM SIGCOMM 2010 Conference}, SIGCOMM '10,
  page 327–338, New York, NY, USA, 2010. Association for Computing Machinery.

\bibitem{7066977}
Stefan Schmid, Chen Avin, Christian Scheideler, Michael Borokhovich, Bernhard
  Haeupler, and Zvi Lotker.
\newblock Splaynet: Towards locally self-adjusting networks.
\newblock {\em IEEE/ACM Transactions on Networking}, 24(3):1421--1433, 2016.

\bibitem{10.1145/1868447.1868455}
Ankit Singla, Atul Singh, Kishore Ramachandran, Lei Xu, and Yueping Zhang.
\newblock Proteus: a topology malleable data center network.
\newblock In {\em Proceedings of the 9th ACM SIGCOMM Workshop on Hot Topics in
  Networks}, Hotnets-IX, New York, NY, USA, 2010. Association for Computing
  Machinery.

\bibitem{278374}
Weitao Wang, Dingming Wu, Sushovan Das, Afsaneh Rahbar, Ang Chen, and
  T.~S.~Eugene Ng.
\newblock {RDC}: {Energy-Efficient} data center network congestion relief with
  topological reconfigurability at the edge.
\newblock In {\em 19th USENIX Symposium on Networked Systems Design and
  Implementation (NSDI 22)}, pages 1267--1288, Renton, WA, April 2022. USENIX
  Association.

\bibitem{10.1145/3351452.3351464}
Klaus-Tycho Foerster and Stefan Schmid.
\newblock Survey of reconfigurable data center networks: Enablers, algorithms,
  complexity.
\newblock {\em SIGACT News}, 50(2):62–79, jul 2019.

\bibitem{179775}
Ankit Singla, P.~Brighten Godfrey, and Alexandra Kolla.
\newblock High throughput data center topology design.
\newblock In {\em 11th USENIX Symposium on Networked Systems Design and
  Implementation (NSDI 14)}, pages 29--41, Seattle, WA, April 2014. USENIX
  Association.

\bibitem{10.1145/331524.331526}
Tom Leighton and Satish Rao.
\newblock Multicommodity max-flow min-cut theorems and their use in designing
  approximation algorithms.
\newblock {\em J. ACM}, 46(6):787–832, nov 1999.

\bibitem{10.1109/SFCS.1988.21958}
T.~Leighton and S.~Rao.
\newblock An approximate max-flow min-cut theorem for uniform multicommodity
  flow problems with applications to approximation algorithms.
\newblock In {\em Proceedings of the 29th Annual Symposium on Foundations of
  Computer Science}, SFCS '88, page 422–431, USA, 1988. IEEE Computer
  Society.

\bibitem{10.1145/2785956.2787510}
Radhika Mittal, Vinh~The Lam, Nandita Dukkipati, Emily Blem, Hassan Wassel,
  Monia Ghobadi, Amin Vahdat, Yaogong Wang, David Wetherall, and David Zats.
\newblock Timely: Rtt-based congestion control for the datacenter.
\newblock In {\em Proceedings of the 2015 ACM Conference on Special Interest
  Group on Data Communication}, SIGCOMM '15, page 537–550, New York, NY, USA,
  2015. Association for Computing Machinery.

\bibitem{10.1145/3341302.3342085}
Yuliang Li, Rui Miao, Hongqiang~Harry Liu, Yan Zhuang, Fei Feng, Lingbo Tang,
  Zheng Cao, Ming Zhang, Frank Kelly, Mohammad Alizadeh, and Minlan Yu.
\newblock Hpcc: High precision congestion control.
\newblock In {\em Proceedings of the ACM Special Interest Group on Data
  Communication}, SIGCOMM '19, page 44–58, New York, NY, USA, 2019.
  Association for Computing Machinery.

\bibitem{278346}
Vamsi Addanki, Oliver Michel, and Stefan Schmid.
\newblock {PowerTCP}: Pushing the performance limits of datacenter networks.
\newblock In {\em 19th USENIX Symposium on Networked Systems Design and
  Implementation (NSDI 22)}, pages 51--70, Renton, WA, April 2022. USENIX
  Association.

\bibitem{10.1145/3387514.3406591}
Gautam Kumar, Nandita Dukkipati, Keon Jang, Hassan M.~G. Wassel, Xian Wu,
  Behnam Montazeri, Yaogong Wang, Kevin Springborn, Christopher Alfeld, Michael
  Ryan, David Wetherall, and Amin Vahdat.
\newblock Swift: Delay is simple and effective for congestion control in the
  datacenter.
\newblock In {\em Proceedings of the Annual Conference of the ACM Special
  Interest Group on Data Communication on the Applications, Technologies,
  Architectures, and Protocols for Computer Communication}, SIGCOMM '20, page
  514–528, New York, NY, USA, 2020. Association for Computing Machinery.

\bibitem{276958}
Prateesh Goyal, Preey Shah, Kevin Zhao, Georgios Nikolaidis, Mohammad Alizadeh,
  and Thomas~E. Anderson.
\newblock Backpressure flow control.
\newblock In {\em 19th USENIX Symposium on Networked Systems Design and
  Implementation (NSDI 22)}, pages 779--805, Renton, WA, April 2022. USENIX
  Association.

\bibitem{10.1145/3387514.3405899}
Ahmed Saeed, Varun Gupta, Prateesh Goyal, Milad Sharif, Rong Pan, Mostafa
  Ammar, Ellen Zegura, Keon Jang, Mohammad Alizadeh, Abdul Kabbani, and Amin
  Vahdat.
\newblock Annulus: A dual congestion control loop for datacenter and wan
  traffic aggregates.
\newblock SIGCOMM '20, page 735–749, New York, NY, USA, 2020. Association for
  Computing Machinery.

\bibitem{10.1145/2018436.2018443}
Christo Wilson, Hitesh Ballani, Thomas Karagiannis, and Ant Rowtron.
\newblock Better never than late: meeting deadlines in datacenter networks.
\newblock In {\em Proceedings of the ACM SIGCOMM 2011 Conference}, SIGCOMM '11,
  page 50–61, New York, NY, USA, 2011. Association for Computing Machinery.

\bibitem{uec}
Ultra ethernet consortium.
\newblock \url{https://ultraethernet.org/}.

\bibitem{abm}
Vamsi Addanki, Maria Apostolaki, Manya Ghobadi, Stefan Schmid, and Laurent
  Vanbever.
\newblock Abm: Active buffer management in datacenters.
\newblock In {\em Proceedings of the ACM SIGCOMM 2022 Conference}, SIGCOMM '22,
  page 36–52, New York, NY, USA, 2022. Association for Computing Machinery.

\bibitem{fab}
Maria Apostolaki, Laurent Vanbever, and Manya Ghobadi.
\newblock Fab: Toward flow-aware buffer sharing on programmable switches.
\newblock In {\em Proceedings of the 2019 Workshop on Buffer Sizing}, BS '19,
  New York, NY, USA, 2020. Association for Computing Machinery.

\bibitem{trafficaware}
Sijiang Huang, Mowei Wang, and Yong Cui.
\newblock Traffic-aware buffer management in shared memory switches.
\newblock {\em IEEE/ACM Transactions on Networking}, 30(6):2559--2573, 2022.

\bibitem{295539}
Vamsi Addanki, Wei Bai, Stefan Schmid, and Maria Apostolaki.
\newblock Reverie: Low pass {Filter-Based} switch buffer sharing for
  datacenters with {RDMA} and {TCP} traffic.
\newblock In {\em 21st USENIX Symposium on Networked Systems Design and
  Implementation (NSDI 24)}, pages 651--668, Santa Clara, CA, April 2024.
  USENIX Association.

\bibitem{10229046}
Hamidreza Almasi, Rohan Vardekar, and Balajee Vamanan.
\newblock Protean: Adaptive management of shared-memory in datacenter switches.
\newblock In {\em IEEE INFOCOM 2023 - IEEE Conference on Computer
  Communications}, pages 1--10, 2023.

\bibitem{295535}
Vamsi Addanki, Maciej Pacut, and Stefan Schmid.
\newblock Credence: Augmenting datacenter switch buffer sharing with {ML}
  predictions.
\newblock In {\em 21st USENIX Symposium on Networked Systems Design and
  Implementation (NSDI 24)}, pages 613--634, Santa Clara, CA, April 2024.
  USENIX Association.

\bibitem{10.1145/2486001.2486031}
Mohammad Alizadeh, Shuang Yang, Milad Sharif, Sachin Katti, Nick McKeown,
  Balaji Prabhakar, and Scott Shenker.
\newblock Pfabric: Minimal near-optimal datacenter transport.
\newblock In {\em Proceedings of the ACM SIGCOMM 2013 Conference on SIGCOMM},
  SIGCOMM '13, page 435–446, New York, NY, USA, 2013. Association for
  Computing Machinery.

\bibitem{259355}
Mohammad Al-Fares, Sivasankar Radhakrishnan, Barath Raghavan, Nelson Huang, and
  Amin Vahdat.
\newblock Hedera: Dynamic flow scheduling for data center networks.
\newblock In {\em 7th USENIX Symposium on Networked Systems Design and
  Implementation (NSDI 10)}, San Jose, CA, April 2010. USENIX Association.

\bibitem{cassini}
Sudarsanan Rajasekaran, Manya Ghobadi, and Aditya Akella.
\newblock Cassini: Network-aware job scheduling in machine learning clusters.
\newblock In {\em 21th USENIX Symposium on Networked Systems Design and
  Implementation (NSDI 24)}, Santa Clara, CA, 2024. USENIX Association.

\bibitem{10.1145/2619239.2626316}
Mohammad Alizadeh, Tom Edsall, Sarang Dharmapurikar, Ramanan Vaidyanathan,
  Kevin Chu, Andy Fingerhut, Vinh~The Lam, Francis Matus, Rong Pan, Navindra
  Yadav, and George Varghese.
\newblock Conga: Distributed congestion-aware load balancing for datacenters.
\newblock In {\em Proceedings of the 2014 ACM Conference on SIGCOMM}, SIGCOMM
  '14, page 503–514, New York, NY, USA, 2014. Association for Computing
  Machinery.

\bibitem{10.1145/2890955.2890968}
Naga Katta, Mukesh Hira, Changhoon Kim, Anirudh Sivaraman, and Jennifer
  Rexford.
\newblock Hula: Scalable load balancing using programmable data planes.
\newblock In {\em Proceedings of the Symposium on SDN Research}, SOSR '16, New
  York, NY, USA, 2016. Association for Computing Machinery.

\bibitem{10.1145/3098822.3098839}
Soudeh Ghorbani, Zibin Yang, P.~Brighten Godfrey, Yashar Ganjali, and Amin
  Firoozshahian.
\newblock Drill: Micro load balancing for low-latency data center networks.
\newblock In {\em Proceedings of the Conference of the ACM Special Interest
  Group on Data Communication}, SIGCOMM '17, page 225–238, New York, NY, USA,
  2017. Association for Computing Machinery.

\bibitem{bonato2025repsrecycledentropypacket}
Tommaso Bonato, Abdul Kabbani, Ahmad Ghalayini, Michael Papamichael, Mohammad
  Dohadwala, Lukas Gianinazzi, Mikhail Khalilov, Elias Achermann, Daniele~De
  Sensi, and Torsten Hoefler.
\newblock Reps: Recycled entropy packet spraying for adaptive load balancing
  and failure mitigation.
\newblock {\em CoRR}, abs/2407.21625, 2025.

\bibitem{addanki2025etherealdivideconquernetwork}
Vamsi Addanki, Prateesh Goyal, Ilias Marinos, and Stefan Schmid.
\newblock Ethereal: Divide and conquer network load balancing in large-scale
  distributed training.
\newblock {\em CoRR}, abs/2407.00550, 2025.

\bibitem{10.1145/3603269.3610862}
Daniel Amir, Tegan Wilson, Vishal Shrivastav, Hakim Weatherspoon, and Robert
  Kleinberg.
\newblock Poster: Scalability and congestion control in oblivious
  reconfigurable networks.
\newblock In {\em Proceedings of the ACM SIGCOMM 2023 Conference}, ACM SIGCOMM
  '23, page 1138–1140, New York, NY, USA, 2023. Association for Computing
  Machinery.

\bibitem{10.1145/3544216.3544254}
Shawn~Shuoshuo Chen, Weiyang Wang, Christopher Canel, Srinivasan Seshan,
  Alex~C. Snoeren, and Peter Steenkiste.
\newblock Time-division tcp for reconfigurable data center networks.
\newblock In {\em Proceedings of the ACM SIGCOMM 2022 Conference}, SIGCOMM '22,
  page 19–35, New York, NY, USA, 2022. Association for Computing Machinery.

\bibitem{246336}
Matthew~K. Mukerjee, Christopher Canel, Weiyang Wang, Daehyeok Kim, Srinivasan
  Seshan, and Alex~C. Snoeren.
\newblock Adapting {TCP} for reconfigurable datacenter networks.
\newblock In {\em 17th USENIX Symposium on Networked Systems Design and
  Implementation (NSDI 20)}, pages 651--666, Santa Clara, CA, February 2020.
  USENIX Association.

\bibitem{10.1145/3651890.3672245}
Jialong Li, Haotian Gong, Federico De~Marchi, Aoyu Gong, Yiming Lei, Wei Bai,
  and Yiting Xia.
\newblock Uniform-cost multi-path routing for reconfigurable data center
  networks.
\newblock In {\em Proceedings of the ACM SIGCOMM 2024 Conference}, ACM SIGCOMM
  '24, page 433–448, New York, NY, USA, 2024. Association for Computing
  Machinery.

\bibitem{286500}
Wei Bai, Shanim~Sainul Abdeen, Ankit Agrawal, Krishan~Kumar Attre, Paramvir
  Bahl, Ameya Bhagat, Gowri Bhaskara, Tanya Brokhman, Lei Cao, Ahmad Cheema,
  Rebecca Chow, Jeff Cohen, Mahmoud Elhaddad, Vivek Ette, Igal Figlin, Daniel
  Firestone, Mathew George, Ilya German, Lakhmeet Ghai, Eric Green, Albert
  Greenberg, Manish Gupta, Randy Haagens, Matthew Hendel, Ridwan Howlader,
  Neetha John, Julia Johnstone, Tom Jolly, Greg Kramer, David Kruse, Ankit
  Kumar, Erica Lan, Ivan Lee, Avi Levy, Marina Lipshteyn, Xin Liu, Chen Liu,
  Guohan Lu, Yuemin Lu, Xiakun Lu, Vadim Makhervaks, Ulad Malashanka, David~A.
  Maltz, Ilias Marinos, Rohan Mehta, Sharda Murthi, Anup Namdhari, Aaron Ogus,
  Jitendra Padhye, Madhav Pandya, Douglas Phillips, Adrian Power, Suraj Puri,
  Shachar Raindel, Jordan Rhee, Anthony Russo, Maneesh Sah, Ali Sheriff, Chris
  Sparacino, Ashutosh Srivastava, Weixiang Sun, Nick Swanson, Fuhou Tian,
  Lukasz Tomczyk, Vamsi Vadlamuri, Alec Wolman, Ying Xie, Joyce Yom, Lihua
  Yuan, Yanzhao Zhang, and Brian Zill.
\newblock Empowering azure storage with {RDMA}.
\newblock In {\em 20th USENIX Symposium on Networked Systems Design and
  Implementation (NSDI 23)}, pages 49--67, Boston, MA, April 2023. USENIX
  Association.

\bibitem{yokar2024fastlinkrecoveryptpsynchronized}
V.~Yokar, A.~Mehrpooya, Y.~Teng, S.~Shen, Z.~Wu, K.~Bardhi, S.~Yan, and
  D.~Simeonidou.
\newblock Fast link recovery via ptp-synchronized nanosecond optical switching,
  2024.

\bibitem{Clark2020SynchronousSC}
Kari~A. Clark, Daniel Cletheroe, Thomas Gerard, Istv{\'a}n Haller, Krzysztof
  Jozwik, Kai Shi, Benn~Charles Thomsen, Hugh Williams, Georgios~S. Zervas,
  Hitesh Ballani, Polina Bayvel, Paolo Costa, and Zhixin Liu.
\newblock Synchronous subnanosecond clock and data recovery for optically
  switched data centres using clock phase caching.
\newblock {\em Nature Electronics}, 3:426--433, 2020.

\bibitem{amplitute}
Thomas Gerard, Kari Clark, Adam Funnell, Kai Shi, Benn Thomsen, Philip Watts,
  Krzysztof Jozwik, Istvan Haller, Hugh Williams, Paolo Costa, and Hitesh
  Ballani.
\newblock Fast and uniform optically-switched data centre networks enabled by
  amplitude caching.
\newblock In {\em 2021 Optical Fiber Communications Conference and Exhibition
  (OFC)}, pages 1--3, 2021.

\bibitem{nvptp}
\url{https://docs.nvidia.com/networking/display/nvidia5ttechnologyusermanualv10}.

\bibitem{sundial}
Yuliang Li, Gautam Kumar, Hema Hariharan, Hassan Wassel, Peter~H. Hochschild,
  Dave Platt, Simon Sabato, Minlan Yu, Nandita Dukkipati, Prashant Chandra, and
  Amin Vahdat.
\newblock Sundial: Fault-tolerant clock synchronization for datacenters.
\newblock In {\em 14th USENIX Symposium on Operating Systems Design and
  Implementation (OSDI 20)}, pages 1171--1186, 2020.

\bibitem{sato}
Ken-Ichi Sato.
\newblock Optical switching will innovate intra data center networks.
\newblock In {\em 2023 Optical Fiber Communications Conference and Exhibition
  (OFC)}, pages 1--40, 2023.

\bibitem{benlee}
Benjamin~G. Lee and Nicolas Dupuis.
\newblock Silicon photonic switch fabrics: Technology and architecture.
\newblock {\em Journal of Lightwave Technology}, 37(1):6--20, 2019.

\bibitem{Raja_2021}
Arslan~Sajid Raja, Sophie Lange, Maxim Karpov, Kai Shi, Xin Fu, Raphael
  Behrendt, Daniel Cletheroe, Anton Lukashchuk, Istvan Haller, Fotini Karinou,
  Benn Thomsen, Krzysztof Jozwik, Junqiu Liu, Paolo Costa, Tobias~Jan
  Kippenberg, and Hitesh Ballani.
\newblock Ultrafast optical circuit switching for data centers using integrated
  soliton microcombs.
\newblock {\em Nature Communications}, 12(1), October 2021.

\bibitem{wu}
Amirmahdi Honardoost, Johannes Henriksson, Kyungmok Kwon, Jianheng Luo, and
  Ming~C. Wu.
\newblock Low-loss wafer-bonded silicon photonic mems switches.
\newblock In {\em 2022 Optical Fiber Communications Conference and Exhibition
  (OFC)}, pages 1--3, 2022.

\end{thebibliography}

\appendix

\section{Frequently Asked Questions}
\label{sec:faq}
\medskip
\noindent
\textbf{Q1:} \textit{How is traffic matrix estimated and at what timescale?}
\smallskip

\begin{figure}[t]
\centering
\includegraphics[width=1\linewidth]{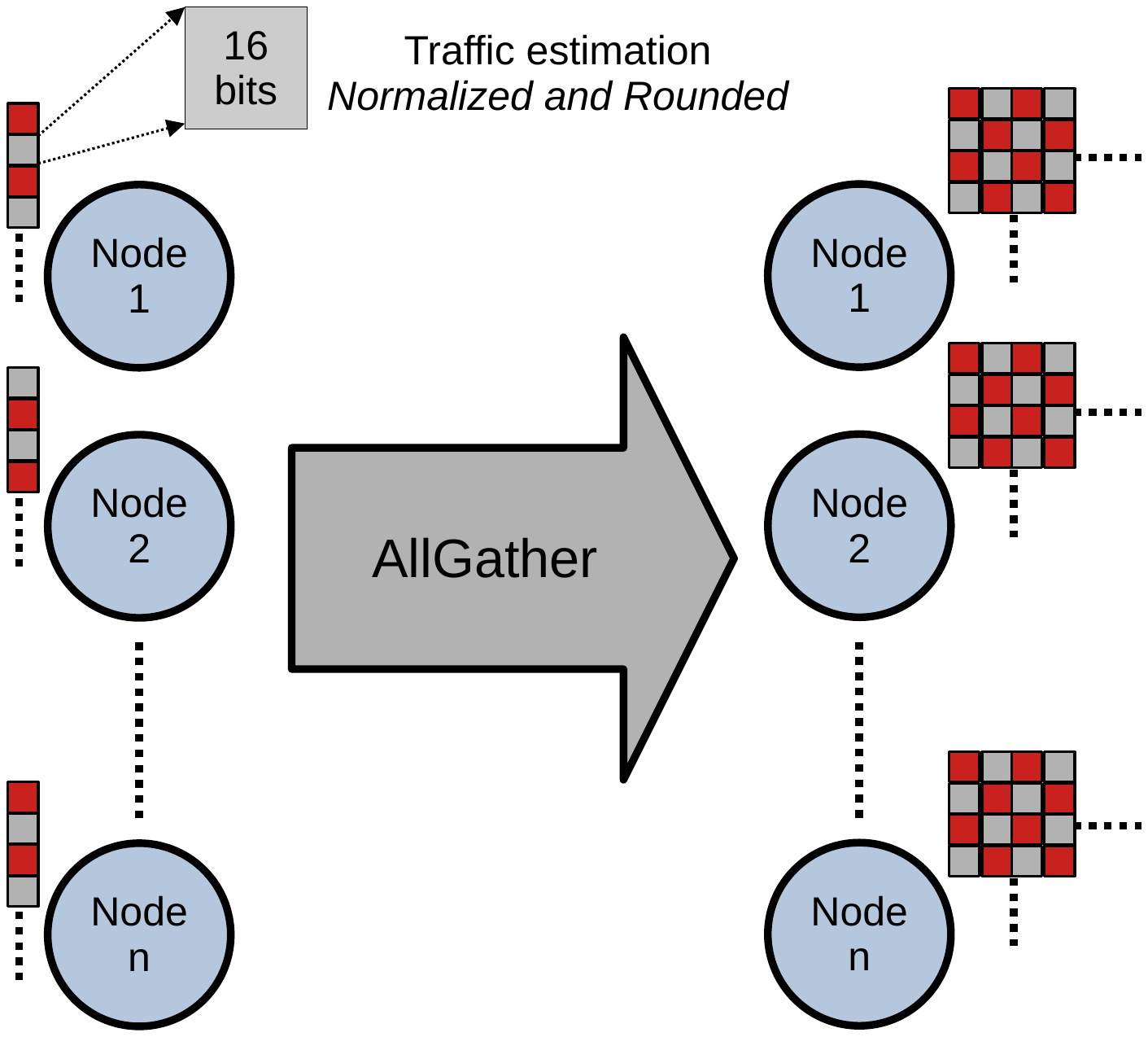}
\caption{Nodes exchange their local traffic estimates during the round-robin phase of \name's schedule by performing an AllGather operation. At the end of the round-robin phase, each node has an overview of the global traffic matrix and performs the schedule computation in a distributed manner.}
\label{fig:estimation-matrix}
\end{figure}

\noindent
\textbf{A1:}
In our current design, each node maintains its own traffic estimation towards every other node in the network. During the round-robin phase of \name's circuit-switching, nodes participate in an AllGather operation pipelined with other data transfers during this phase. The AllGather operation runs as follows:
\begin{itemize}[label=\small{\textcolor{myred}{$\blacksquare$}}]
    \item Each node maintains a local array of traffic estimates (in bits) towards every other node in the network i.e., a single row of the global traffic matrix.

    \item Each node retrieves its local array of outgoing traffic estimates corresponding to all the destination nodes in the network and transforms the array in two steps. First, it performs normalization by multiplying each entry in the array by $\frac{k-1}{k}\cdot\frac{1}{c\cdot \Delta}$, where $c$ is the physical link capacity, $k$ is the number of phases in a period (a parameter to \name) and $\Delta$ is the timeslot duration\footnote{$c\cdot \Delta$ corresponds to the total amount of bits that each node can transmit in a single timeslot, which is known in advance and remains constant i.e., based on the physical link capacity and the timeslot duration of the network.}. Second, it rounds down the normalized entries.
    Importantly, the left over non-zero entries at each node are bounded by the total number of nodes $n$ in the network. Further, each entry's value and the sum of all entries in the array are bounded by $k\cdot n$. We allocate $16$ bits to represent each entry, supporting up to $65536$ values corresponding to up to $n=21845$ number of ToRs in the network. 

    \item During the round-robin phase, each node sends its array of traffic estimations to its direct neighbor in every timeslot --- essentially an AllGather operation.

    \item At the end of the round-robin phase, every node has an overview of the global traffic matrix (normalized and rounded) as shown in Figure~\ref{fig:estimation-matrix}.
\end{itemize}

The amount of data transmitted by each node during a single timeslot in the round-robin phase is limited by the number of nodes $n$. For example, even with a large-scale network of $n=256$ ToRs, the total data transmitted by a node in a single timeslot amounts to $256 \times 16$ bits. With a link capacity of $800$ Gbps, this transmission requires only $\approx 41$ nanoseconds. Encouragingly, port bandwidth tends to increase over time, further reducing the time required to transmit traffic estimates. This trend suggests that even a conservative estimate of $41$ nanoseconds for the minimum timeslot duration for circuit-switching is sufficient for \name to efficiently perform the AllGather operation and construct the traffic matrix at each node.

\bigskip
\noindent
\textbf{Q2:} \textit{How do nodes estimate their local traffic array?}
\smallskip

\noindent
\textbf{A2:} Each source ToR switch maintains per-destination (ToR) virtual output queues (VOQs). Each entry in the local traffic estimate array at each ToR represents the total number of bits received by the corresponding VOQ. These counters are reset at the end of each round-robin phase, once the transmission of traffic estimates is complete.

Looking ahead, we aim to explore the feasibility of servers reporting local traffic estimates directly to their upstream ToR switches. Server NICs, particularly in RDMA environments, provide more accurate traffic estimates. Specifically, NICs can scan RDMA queue pairs to calculate the total number of bytes specified by the outstanding WQEs across all active queue pairs. This practice, widely used in production for congestion control, telemetry, and buffer management~\cite{286500}, offers an intriguing advantage. Notably, WQEs do not represent bytes waiting for immediate transmission but rather the \emph{outstanding} bytes the NIC \emph{intends} to transmit in the future. This forward-looking information provides a more accurate traffic estimate compared to relying solely on VOQs at the ToR switch. We leave a deeper investigation of this approach for future work.

\begin{figure}[t]
\centering
\includegraphics[width=1\linewidth]{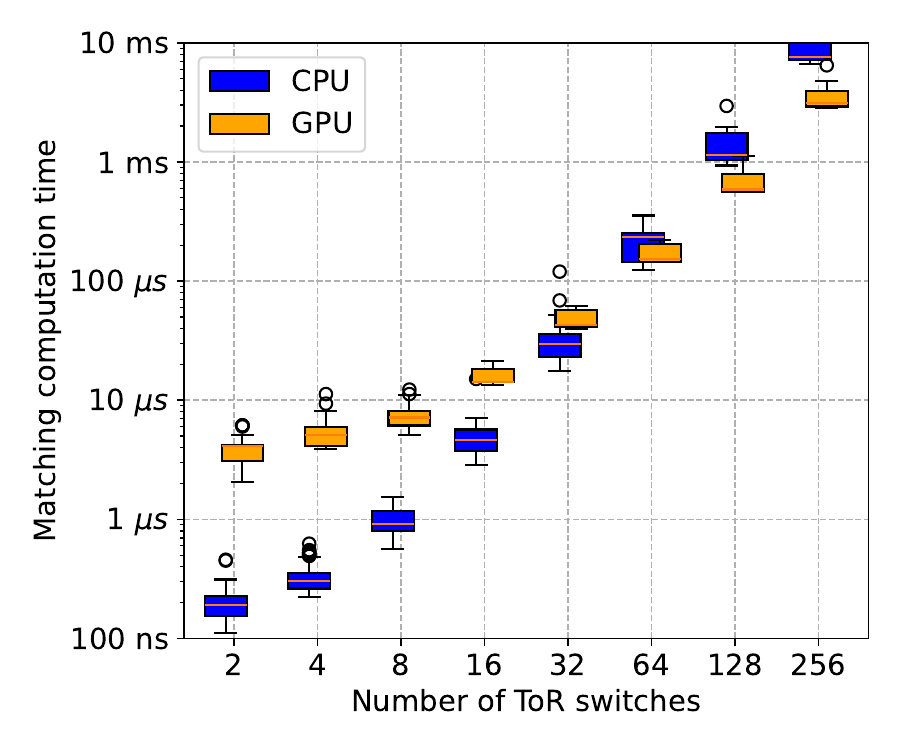}
\caption{The absolute amount of time required for computing the circuit-switching schedule of \name is within microsecond scale for moderate network sizes e.g., with $32$ nodes (ToRs) and $32$ servers per ToR, \name incurs $59\mu s$ of computation time on average for a network size of $1024$ servers.}
\label{fig:matching-times}
\end{figure}

\bigskip
\noindent
\textbf{Q3:} \textit{What is the absolute time required to compute \name's schedule?}

\noindent
\textbf{A3:} Figure~\ref{fig:matching-times} shows the time taken to compute \name's schedule for a range of network sizes between $2$ to $256$ ToR switches. We wrote a CUDA program that takes a doubly stochastic matrix as input and computes the matching decomposition. We tested the computation time on an NVIDIA RTX A2000 GPU. 

We observe that the time taken to compute the matching decomposition is within microsecond scale for moderate network sizes. 
In the following, we consider $32$ servers per ToR switch. For instance, Broadcom Tomahawk $5$ supports $64$ ports of $800$Gbps each. With a similar ToR switch radix, $32$ servers per ToR is a reasonable assumption. Figure~\ref{fig:matching-times} shows that for $32$ ToRs (i.e, $1024$ servers) in the network, \name only incurs $59\mu s$ on average for computing the switching schedule. As the network size increases, the time required for computing the switching schedule increases significantly beyond acceptable limits. In this regard, we currently target up to $32$ ToR network size if the optical switches are deployed at the spine layer. For larger network sizes, \name may be best suited for circuit-switching at the core, i.e., between spine (or aggregation) and the core switches, similar to Google's Jupiter~\cite{10.1145/3544216.3544265}. Better algorithms for matching decomposition, as well as hardware acceleration, could further reduce the computation time and extend the network size that \name can support.

\bigskip
\noindent
\textbf{Q4:} \textit{How frequently is the new schedule computed and how does it impact the freshness of traffic estimates?}
\smallskip

\noindent
\textbf{A4:} \name performs the traffic matrix AllGather operation in \emph{every} period during the round-robin phase as described in (A1). However, the computations for the matching decomposition are performed asynchronously. 

\begin{itemize}[label=\small{\textcolor{myred}{$\blacksquare$}}]

    \item \textbf{Frequency of computations:} Given that the computation happens asynchronously, the frequency of the computations is dictated by the computation time. As shown in Figure~\ref{fig:matching-times}, the computation time is within microsecond scale for moderate network sizes \eg $59\mu s$ for a network size of $32$ ToR switches. This allows for frequent updates to the schedule, ensuring that the network adapts to changing traffic patterns even at microsecond timescales. \name updates its schedule only after the computation results are ready and does not block the circuits and communication. We enforce that such an update only happens at the end of a round-robin phase, allowing new computations to account for the most recent traffic estimates gathered by the previous round-robin phase.

    \item \textbf{Time synchronization and Link recovery:} Dynamic optical circuit-switched networks critically rely on time synchronization, including a vast majority of the designs in the literature~\cite{10.1145/3098822.3098838,10.1145/3387514.3406221,opera,10.1145/3651890.3672273,10.1145/3651890.3672248}. In comparison to prior works, \name additionally requires that any updates to the circuit-switching schedule happen in a precisely synchronized manner. This synchronization is crucial to prevent inconsistencies in the schedule and to avoid undesirable optical collisions.
    Accurate time synchronization has been demonstrated experimentally in the literature~\cite{10.1145/3387514.3406221, yokar2024fastlinkrecoveryptpsynchronized}. In addition, transmission of packets to the optical network needs to be time-synchronous, to ensure data is transmitted when the appropriate optical channel is setup~\cite{corundum}. Link bring-up (clock/data recovery CDR) overheads after each reconfiguration of the network needs to be minimized and relevant approaches have been reported in the literature ~\cite{10.1145/3387514.3406221, yokar2024fastlinkrecoveryptpsynchronized,Clark2020SynchronousSC,10.1145/3651890.3672273, amplitute}. Notably, recent developments of commercial solutions ~\cite{nvptp, sundial} can be leveraged to address these challenges. We leave it for future work to explore the synchronization and CDR mechanisms in more detail.

    \item \textbf{Freshness of traffic estimates:} The AllGather operation collects the traffic estimates during every robin-robin phase but computations are performed less frequently in an asynchronous manner.  We apply exponential weighted moving averages for each entry in the local traffic estimates at each node. This preserves the freshness of traffic estimates, as well as, takes into account the historical estimates received between two computations.
\end{itemize}

\bigskip
\noindent
\textbf{Q5:} \textit{Any variable-duration schedule obtained from BvN decomposition can be converted to a periodic fixed-duration schedule by time quantization. Why is \name's periodic schedule different?}
\smallskip

\noindent
\textbf{A5:} It is certainly true that any BvN schedule can be converted to a periodic fixed-duration schedule by time quantization. However, such an approach has two critical issues in terms of performance under realistic reconfiguration delays, in contrast to the ideal performance as indicated by Theorem~\ref{th:throughput-ideal}:

\begin{itemize}[label=\small{\textcolor{myred}{$\blacksquare$}}]

\item \textbf{Schedule length:} The length of the schedule obtained from BvN decomposition can be up to $n^2 -2$ matchings, in comparison to just $3\cdot n$ matchings (for $k=3$) in \name. This results in a significantly longer schedule length for large networks. For instance with $n=32$, a BvN schedule could produce up to $1024$ matchings, whereas, \name consistently uses only a period length of $96$ which scales down proportionately with the number of physical links (degree) at each ToR switch.

\item \textbf{Throughput:} More critically, time quantization of a BvN schedule can result in significantly lower throughput due to the reduced circuit duty cycle. For instance, consider $4$ matchings produced by BvN for a traffic matrix $\mathcal{M}$, i.e., $\mathcal{M} = \lambda_1 \cdot P_1 + \lambda_2 \cdot P_2 + \lambda_3 \cdot P_3 + \lambda_4 \cdot P_4$. To achieve full throughput in an ideal scenario with zero reconfiguration delay (see Theorem~\ref{th:throughput-ideal}), it is necessary to spend $t_1 = \frac{\lambda_1}{\sum_i \lambda}$ fraction of time executing the matching $P_1$, $t_2 = \frac{\lambda_2}{\sum_i \lambda}$ fraction of time executing the matching $P_2$, and so on.

For clarity, we first define \emph{fixed-duration} and \emph{variable-duration} circuit-switching.

\begin{definition}[Fixed-duration circuit-switching]\label{def:fixed-duration-schedule}
Any circuit-switching schedule is a sequence of matchings $P = \langle P_1,\ P_2,\ \dots \rangle $. A fixed-duration circuit-switching schedule specifies that every matching is executed for a fixed duration $\Delta$ for all matchings $P_i \in P$.
\end{definition}

For example, in a $4$-node network with nodes labeled $A$, $B$, $C$, and $D$, suppose we desire a circuit $A \rightarrow C$ for a duration of $2$ timeslots and a circuit $B \rightarrow D$ for a duration of $3$ timeslots. This setup represents a variable-duration circuit-switching schedule. However, it is possible to construct a series of matchings with fixed durations: $P_1 = \langle A \rightarrow C,\ B \rightarrow D,\ D \rightarrow A,\ C \rightarrow B \rangle$, $P_2 = \langle A \rightarrow C,\ B \rightarrow D,\ D \rightarrow A,\ C \rightarrow B \rangle$, $P_3 = \langle A \rightarrow B,\ B \rightarrow D,\ D \rightarrow C,\ C \rightarrow A \rangle$. 

\medskip
This schedule provides the desired circuit durations---a circuit between $A \rightarrow C$ for $2$ timeslots and a circuit between $B \rightarrow D$ for $3$ timeslots. However, this schedule is \textbf{\textit{not} a variable-duration} schedule, as each matching requires a fixed duration. As long as each \emph{matching} is assigned a fixed duration, the schedule is considered a fixed-duration schedule.

\begin{definition}[Variable-duration circuit-switching]
\label{def:variable-duration-schedule}
Any circuit-switching schedule is a sequence of matchings $P = \langle P_1,\ P_2,\ \dots \rangle $. A variable-duration circuit-switching schedule specifies the execution time $t_i$ for each matching $P_i$ for all matchings $P_i \in P$. Importantly, the time duration for each matching is not a fixed value and can differ across matchings.
\end{definition}

Based on Definition~\ref{def:fixed-duration-schedule} and Definition~\ref{def:variable-duration-schedule}, we can formally articulate the challenge of converting BvN schedules into fixed-duration schedules. BvN schedules are inherently variable-duration schedules, as the execution time $t_i = \frac{\lambda_i}{\sum_i \lambda_i}$ for each matching $P_i$ is not guaranteed to be uniform across all matchings.

\medskip
Consider the example described above with four matchings, each requiring durations $t_1$, $t_2$, $t_3$, and $t_4$, respectively. For simplicity, assume $t_1 > t_2 > t_3 > t_4$. To convert this schedule into a fixed-duration schedule, we must select a timeslot duration $\Delta$ such that $t_1, t_2, t_3, t_4 \ge \Delta$. Suppose we choose $\Delta = t_4$. By time quantization, this would require repeating the first matching $P_1$ for $\frac{t_1}{t_4}$ times, the second matching $P_2$ for $\frac{t_2}{t_4}$ times, and so on.

\medskip
The critical challenge in such quantization is that $\frac{t_1}{t_4}$ can be arbitrarily large, as the schedules generated by BvN decomposition do not guarantee a minimum value for the coefficients $\lambda$, which can lead to enormously long schedules.
Unfortunately, there is no straightforward solution in the literature to quantize BvN schedules without compromising the schedule length and duty cycle. We view \name's matrix rounding approach as effectively mimicking such a quantization process, providing a fixed-duration schedule while achieving provably high throughput.

\end{itemize}

\bigskip
\noindent
\textbf{Q6:} \textit{What are the implementation alternatives for fast-switching fabrics?}
\smallskip

\noindent
\textbf{A6:} There are two primary approaches to achieving fast optical switching~\cite{sato}: fast spatial optical switches ~\cite{benlee} and wavelength switching, that can be achieved with the combination of fast tunable lasers with wavelength selective elements like arrayed waveguide grating routers (AWGRs) ~\cite{10.1145/3387514.3406221}. 

Tunable lasers combined with AWGRs enable wavelength-based routing with high speed. Recent advancements in laser designs~\cite{Raja_2021}—especially disaggregated architectures that separate wavelength generation from selection—have reduced tuning latencies to nanosecond levels. In addition, AWGRs are passive and robust, requiring no mechanical components or frequent upgrades. 
Nevertheless, the widespread adoption of this approach has been hindered by practical implementation tradeoffs. Cost-effective tunable lasers with fast tuning capabilities are not yet widely available and consume more power than their fixed-wavelength counterparts. In AWGR-based systems scalability to large ports counts is limited by the number of wavelengths available in the system.

Spatial optical switches are predominantly based on either Mach–Zehnder interferometers (MZI) or Microelectrical Mechanical Systems (MEMS) on silicon photonic platforms. Electro-optic MZI-based switches \cite{benlee, sato} have been shown to achieve ns scale switching. However, they exhibit high insertion loss and increased crosstalk that limits their scalability. Silicon photonic MEMS switches ~\cite{wu}  have been shown to achieve switching in the order of 1-10 microseconds and can reach high radix, with lower insertion loss and better crosstalk performance. Therefore, despite their somewhat slower switching speed and reliance on custom fabrication process, silicon photonic MEMS switches currently represent promising candidates for scalable fast optical switching fabrics. We leave if for future work to explore the hardware implementation of the switching fabric.

\medskip
\noindent\textbf{Q7:}\textit{ Is it computationally feasible to derive optimal traffic-aware periodic schedules for large topologies?}

\medskip
\noindent\textbf{A7:}
The underlying problem is to construct an optimal emulated topology within a degree constraint. Appendix~\ref{sec:linear-program} presents the linear program formulation. However, in our initial experiments, the solver (Gurobi~\cite{gurobi}) did not terminate after $30$ minutes even for a $16$ node topology. Our approach addresses this challenge with an novel algorithm that can quickly derive near-optimal schedules for large topologies and demand matrices.

\medskip
\noindent\textbf{Q8:} \textit{Given a degree constraint, what is the best  topology that maximizes throughput for a given traffic matrix?}

\medskip
\noindent\textbf{A8:}
Answering this question not only allows us to then derive a periodic schedule that emulates an optimal topology but also provides insights into more constrained networks with slow reconfigurations such as those with patch-panels~\cite{285119}. While some recent works focus on specific communication patterns in distributed training under a domain-specific cost model~\cite{efficientdirectnsdi2025}, the throughput-optimization problem largely remains open for general communication patterns. Our roadmap to solve this problem builds upon our observations in \S\ref{sec:traffic-aware}. Specifically, building upon Theorem~\ref{th:throughput-integer}, our main intuition is to serve the bulk portion of the traffic matrix in a traffic-aware manner using a subset of the available links within the degree constraints and tackle the residual demand with a traffic-oblivious topology, while guaranteeing high throughput.
We address this question for specific degree constraints relevant for periodic networks but under general demand matrices within the hose model, providing a first step towards a general solution (see \S\ref{sec:properties}).

\medskip
\noindent\textbf{Q9:} \textit{Is it fundamentally feasible to achieve high throughput for any traffic matrix using only single-hop routing?}

\medskip
\noindent\textbf{A9:}
Under constrained length of the schedule (and degree), answering this question not only requires finding optimal topology under an ``ideal routing'' obtained from solving the concurrent flow problem, we further need to restrict the paths to direct communication. Our approach is to construct \emph{short} periodic schedules that provide bandwidth between communicating pairs such that the gap between bandwidth and demand between any pair is bounded by a certain ratio, ensuring high throughput with single-hop routing (see \S\ref{sec:properties}).

\section{Throughput of \\traffic-aware Networks}
\label{app:throughput}

\idealThroughputTheorem*

\begin{proof}
Within the hose model set of demand matrices, we consider saturated demand matrices \ie the sum of every row (column) equals the outgoing (incoming) capacity of each node. If a topology can achieve throughput $\theta$ for all saturated demand matrices, then the topology can achieve throughput $\theta$ for any traffic matrix~\cite{10.1145/3452296.3472913}. Given that saturated demand matrices are doubly stochastic, we first decompose the matrix using Birkhoff–von Neumann (BvN) decomposition technique~\cite{birkhoff1946three} into $k$ permutation matrices, where $k$ can be up to $n^2$. Let $\mathcal{M}$ be any saturated traffic matrix, where the sum of every row and column is $c\cdot u$ (total capacity of each node). Let the corresponding BvN decomposition be $\mathcal{M} = \lambda_1 \cdot P_1 + \lambda_2\cdot P_2 ... + \lambda_k P_k$, where $P_i$ is a permutation matrix and the coefficients $\lambda$ are such that $\sum_{i=1}^k \lambda = c\cdot u$.
Using this decomposition, we configure the topology such that each permutation $P_i$ is executed using full node capacity $c\cdot u$ for $\frac{\lambda_i}{c\cdot u}\cdot \Delta$ units of time over a period of one unit of time $\Delta$. Over $\Delta$ amount of time, $\lambda_i\cdot P_i$ portion of the traffic matrix generates $\lambda_i\cdot P_i\cdot \Delta$ demand in volume. As a result, during $\frac{\lambda_i}{c\cdot u}\cdot \Delta$ amount of time, by executing the corresponding permutation $P_i$ using full capacity $c\cdot u$, the topology can fully satisfy $\lambda_i\cdot P_i$ portion of the traffic matrix.
As a result, the topology can fully satisfy the traffic matrix $\mathcal{M}$ over each period of one unit of time $\Delta$ and achieves full throughput. 
\end{proof}

\begin{figure*}
\centering
\begin{subfigure}{1\linewidth}
\centering
\includegraphics[width=0.8\linewidth]{plots/throughput-16-legend.pdf}
\end{subfigure}
\begin{subfigure}{1\linewidth}
\centering
\includegraphics[width=0.9\linewidth]{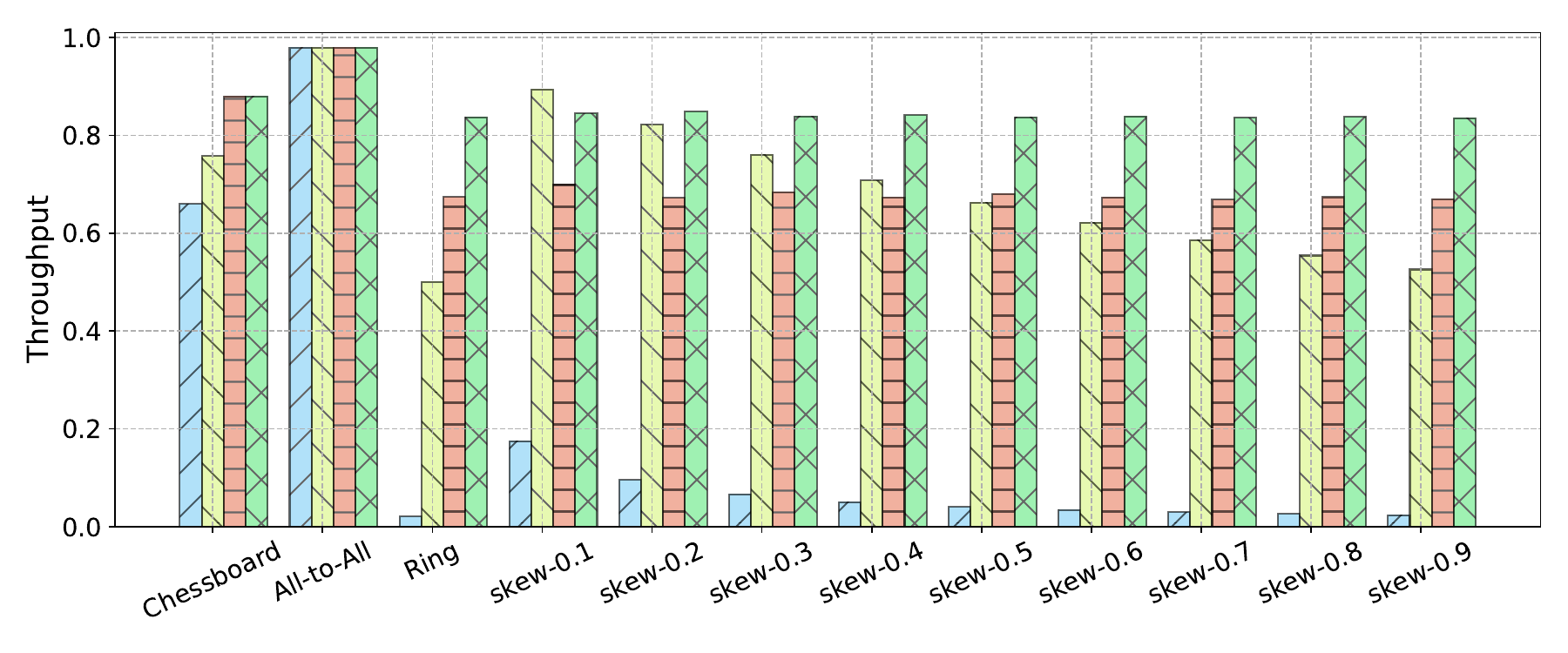}
\end{subfigure}
\caption{Throughput of oblivious periodic network and \name for a $48$ node topology. \name achieves higher throughput compared to oblivious networks, even with single-hop direct communication.}
\label{fig:throughput-48}
\end{figure*}

\intergerMatrixTheorem*

\begin{proof}
A traffic matrix $\mathcal{M}$ in the hose model has the property that the sum of every row and column is at most $\hat{d}$, where $\hat{d}$ is the degree (physical links). Further, we have that every entry in the matrix is an integer multiple of $c\cdot \frac{1}{\Gamma}$, where $c$ is the link capacity and $\Gamma$ is the period of the periodic schedule. We assume, without loss of generality, that the capacity of each physical link is $1$.
The emulated multigraph $G$ has a degree of $\Gamma \cdot \hat{d}$, and each link has capacity $\frac{1}{\Gamma}$.
We now upscale both the traffic matrix $\mathcal{M}$ and each link capacity of the emulated multigraph by $\Gamma$. Thus, it is equivalent to find the throughput of the emulated multigraph $G$ with degree $\Gamma \cdot \hat{d}$ and each link having capacity $1$, under the scaled traffic matrix $\mathcal{M}^{\prime} = \Gamma \cdot \mathcal{M}$, where the sum of each row and column is at most $\Gamma\cdot \hat{d}$.
Since every entry in the matrix $\mathcal{M}$ is an integer multiple of $\frac{1}{\Gamma}$, the scaled traffic matrix $\mathcal{M}^{\prime}$ is an integer matrix with sum of every row and column at most $\Gamma\cdot \hat{d}$. Further, we have a degree of $\Gamma \cdot \hat{d}$ with link capacity of $1$. It is now straight-forward that constructing a graph by adding links between each pair based on the value of the demand in the scaled matrix $\mathcal{M}^\prime$ can fully satisfy the demand, and this requires only single-hop routing.
\end{proof}

\section{Linear Program Formulation}
\label{sec:linear-program}
\textit{We emphasize that the following linear program formulation \textbf{is not related to \name's design}.} Instead, the goal of this section is to formulate the underlying problem of finding the optimal emulated graph for a given traffic matrix under a degree constraint. This formulation serves as a theoretical exercise to understand the complexity of the problem, which motivates the development of efficient algorithms like \name.

Throughput maximization is a variant of multi-commodity maximum flow problem, commonly referred as maximum concurrent flow problem~\cite{10.1145/77600.77620}. In the case of traffic-aware periodic networks, our goal is to find the best emulated graph. In the following we present an integer linear program formulation. Given a network of $n$ nodes, each with $\hat{d}$ physical links (incoming and outgoing), the LP takes traffic matrix $\mathcal{M}$ and the length of the desired schedule $\Gamma$ as input. The capacity of the physical links is denoted by $c$.
The LP has to find the number of links $\hat{c}^{i,j}$ between each node pair $(i,j)$.
We set the link capacities to $1$ and interpret $\hat{c}^{i,j}$ as the capacity between $i,j$. We use $f_{i,j}^{s,d}$ to refer to the flow on edge $(i,j)$ corresponding to $(s,d)$ demand. Our objective is to maximize throughput $\theta$ such that the scaled traffic matrix $\theta\cdot \mathcal{M}$ satisfies source-destination demands, flow conservation and capacity constraints. We consider a degree constraint $\hat{d}$ for each node. Consequently, the demand matrices of interest are those with the sum of each row and column limited to $\hat{d}$. 

\medskip
\noindent\textbf{Input:} 
\begin{align*}
\text{traffic matrix} & \quad \mathcal{M} = \{m_{s,d} \mid s\in V,\ d\in V\}  \\
\text{In-out degree} & \quad \hat{d}
\end{align*}

\noindent \textbf{Objective Function:}
\begin{align*}
\text{Maximize} \quad & \theta
\end{align*}

\noindent \textbf{Subject to the constraints:}
\begin{align*}
\text{Source demand:} & \sum_{i\in V\backslash \{s\}} f^{s,d}_{s,i} \ge \theta \cdot m_{s,d} \\
& \quad\quad\quad\quad\quad\quad\quad \forall s\in V,\ \forall d \in V \\
\text{Destination demand:} & \sum_{i\in V\backslash \{d\}} f^{s,d}_{i,d} \ge \theta \cdot m_{s,d}\\ 
& \quad\quad\quad\quad\quad\quad\quad \forall s\in V,\ \forall d \in V  \\
\text{Flow conservation:} & \sum_{i\in V\backslash\{j\}} f^{s,d}_{i,j} - \sum_{k\in V\backslash\{j\}} f^{s,d}_{j,k} = 0  \\
& \quad\quad\quad\quad\quad\quad\quad\quad\forall j \in V\backslash\{s,d\}\\
& \quad\quad\quad\quad\quad\quad\quad\forall s\in V,\ \forall d \in V\\
\end{align*}
\begin{align*}
\text{Capacity constraints:} & \sum_{s\in V} \sum_{d\in V} f^{s,d}_{i,j} \le \hat{c}^{i,j} \\
& \quad\quad\quad\quad\quad\quad\quad\forall i\in V,\ \forall j \in V \\
\text{In-degree constraints:} & \sum_{s\in V} c^{s,d} \le \hat{d} \\
& \quad\quad\quad\quad\quad\quad\quad\forall d\in V\\
\text{Out-degree constraints:} & \sum_{s\in V} c^{s,d} \le \hat{d} \\
& \quad\quad\quad\quad\quad\quad\quad\forall d\in V
\end{align*}
\noindent \textbf{Variables:}
\begin{align*}
\text{Flow: } & f_{i,j}^{s,d} \ge 0 \ , \ f_{i,j}^{s,d} \in \mathbb{R} \\
& \quad\quad\quad\quad\quad\forall i\in V, j\in V, s\in V,\ \forall d \in V\\
\text{Throughput: } & \theta\ge 0 \ , \ \theta \in \mathbb{R} \\
\text{Number of links: } & \hat{c}^{i,j} \ge 0 \ , \ \hat{c}^{i,j} \in \mathbb{Z} \\
\end{align*}

\section{Additional Results}\label{app:additional-results}

Figure~\ref{fig:throughput-48} shows our throughput results for a network size of $48$ nodes. These results follow the similar observations as the $16$ node network used in \S\ref{sec:gurobi}.

\label{LastPage}
\end{document}